\documentclass[12pt]{article}
\pdfoutput=1
\usepackage{latexsym}
\usepackage{amssymb,amsfonts,amsmath}
\usepackage{graphicx} 
\usepackage{indentfirst}
\usepackage{bbm}
\usepackage{amssymb}
\usepackage{verbatim}
\usepackage{amsmath, amsthm,amssymb}
\usepackage{mathrsfs}
\usepackage{amsfonts}
\usepackage{dsfont}
\usepackage{csquotes}
\usepackage{hyperref}
\usepackage{url}
\usepackage{xcolor}
\usepackage{xpatch}
\usepackage[english]{babel}

\usepackage[style=phys, articletitle=true, biblabel=brackets, backend=bibtex,
chaptertitle=false, pageranges=false, eprint=true,%
natbib=true, maxbibnames=10, giveninits=true, sorting=none]{biblatex} 

\DeclareFieldFormat[article]{journaltitle}{\mkbibitalic{#1\isdot}} 
\makeatletter
\newrobustcmd{\mkbibfixedbrackets}[1]{%
	\begingroup
	\blx@blxinit
	\blx@setsfcodes
	\bibleftbracket#1\bibrightbracket
	\endgroup}

\renewbibmacro*{doi+eprint+url}{%
	\iftoggle{bbx:doi}
	{\printfield{doi}}
	{}%
	\ifboolexpr{test {\iffieldequalstr{eprinttype}{arxiv}} or test {\iffieldequalstr{eprinttype}{arXiv}}}
	{\setunit{\addspace}\newblock}
	{\newunit\newblock}%
	\iftoggle{bbx:eprint}
	{\usebibmacro{eprint}}
	{}%
	\newunit\newblock
	\iftoggle{bbx:url}
	{\usebibmacro{url+urldate}}
	{}}

\xpatchbibdriver{online}
{\newunit\newblock
	\iftoggle{bbx:eprint}
	{\usebibmacro{eprint}}
	{}}
{\ifboolexpr{test {\iffieldequalstr{eprinttype}{arxiv}} or test {\iffieldequalstr{eprinttype}{arXiv}}}
	{\setunit{\addspace}\newblock}
	{\newunit\newblock}%
	\iftoggle{bbx:eprint}
	{\usebibmacro{eprint}}
	{}}
{}{}

\DeclareFieldFormat{eprint:arxiv}{%
	\mkbibbrackets{%
		\ifhyperref
		{\href{http://arxiv.org/\abx@arxivpath/#1}{%
				arXiv\addcolon
				\nolinkurl{#1}%
				\iffieldundef{eprintclass}
				{}
				{\addspace\UrlFont{\mkbibfixedbrackets{\thefield{eprintclass}}}}}}
		{arXiv\addcolon
			\nolinkurl{#1}%
			\iffieldundef{eprintclass}
			{}
			{\addspace\UrlFont{\mkbibfixedbrackets{\thefield{eprintclass}}}}}}}
\makeatother

\parskip=\medskipamount

\newcommand{\cA}{{\cal A}}
\newcommand{\cB}{{\cal B}}
\newcommand{\cC}{{\cal C}}
\newcommand{\cD}{{\cal D}}

\newcommand{\cF}{{\cal F}}

\newcommand{\cH}{{\cal H}}
\newcommand{\cI}{{\cal I}}
\newcommand{\cJ}{{\cal J}}

\newcommand{\cN}{{\cal N}}

\newcommand{\cQ}{{\cal Q}}

\newcommand{\cZ}{{\cal Z}}

\def\a{\alpha}

\def\b{\beta}

\def\d{\delta}
\def\e{\epsilon}
\def\f{\phi}
\def\g{\gamma}

\def\l{\lambda}
\def\m{\mu}
\def\n{\nu}

\def\q{\theta}
\def\r{\rho}
\def\s{\sigma}
\def\t{\tau}

\def\x{\xi}

\def\F{\Phi}
\def\J{\Psi}
\def\L{\Lambda}

\def\P{\Pi}
\def\Q{\Theta}

\newcommand{\ve}{\varepsilon}                            

\newcommand{\pa}{\partial}                           

\newcommand{\abar}{\bar{a}}
\newcommand{\bbar}{\bar{b}}
\newcommand{\cbar}{\bar{c}}
\newcommand{\be}{\begin{equation}}
\newcommand{\ee}{\end{equation}}
\newcommand{\bea}{\begin{eqnarray}}
\newcommand{\eea}{\end{eqnarray}}

\newcommand{\ba}{\begin{array}}
\newcommand{\ea}{\end{array}}

\def\double #1{#1{\hbox{\kern-2pt $#1$}}}

\newcommand{\sSO}{\mathsf{SO}}
\newcommand{\sO}{\mathsf{O}}

\newcommand{\bsubeq}{\begin{subequations}}
\newcommand{\esubeq}{\end{subequations}}

\numberwithin{equation}{section}

\begin{document}

\begin{titlepage}
\begin{flushright}
April, 2021
\end{flushright}
\vspace{2mm}

\begin{center}
\Large \bf Mixed three-point functions of conserved currents in three-dimensional superconformal field theory  \\
\end{center}

\begin{center}
{\bf
Evgeny I. Buchbinder and Benjamin J. Stone}

{\footnotesize{
{\it School of Physics M013, The University of Western Australia\\
35 Stirling Highway, Crawley W.A. 6009, Australia}} ~\\
}
\end{center}
\begin{center}
\texttt{Email: evgeny.buchbinder@uwa.edu.au, \\ benjamin.stone@research.uwa.edu.au}
\end{center}

\vspace{4mm}

\begin{abstract}
\baselineskip=14pt

We consider mixed three-point correlation functions of the supercurrent and flavour current in three-dimensional
$1 \leq \cN \leq 4$ superconformal field theories. Our method is based on the decomposition of the relevant tensors into irreducible components
to guarantee that all possible tensor structures are systematically taken into account. 
We show that only parity even structures appear in the correlation functions. 
In addition to the previous results obtained in \href{https://doi.org/10.1007/JHEP06(2015)138}{arXiv:1503.04961},
it follows that supersymmetry forbids parity odd structures in three-point functions involving the supercurrent and flavour current multiplets.
\end{abstract}
\end{titlepage}

\newpage
\renewcommand{\thefootnote}{\arabic{footnote}}
\setcounter{footnote}{0}

\tableofcontents
\vspace{1cm}
\bigskip\hrule

\section{Introduction}\label{section1}

It is a well-known property of conformal field theories that the functional form of two- and three-point functions of conserved currents such as the energy-momentum tensor and vector current are fixed up to finitely many parameters. In \cite{Osborn:1993cr,Erdmenger:1996yc} a systematic formalism was developed to construct 
two- and three-point functions of primary operators in diverse dimensions. The method was based on properly imposing the relevant symmetries arising from 
scale transformations and permutations of points as well as the conservation laws for the conserved currents, (see also refs.~\cite{Polyakov:1970xd,Schreier:1971um, Ferrara:1972cq,  Ferrara:1973yt, Koller:1974ut, Mack:1976pa, Fradkin:1978pp, Stanev:1988ft} 
for earlier work). 
More recently it was shown in~\cite{Giombi:2011rz} that a peculiar feature of three-dimensional 
(and perhaps in general, odd-dimensional) conformal field theories is the appearance of parity violating contributions in three-point functions of conserved currents. 
These structures were overlooked in the original study by Osborn and Petkou \cite{Osborn:1993cr} (also \cite{Erdmenger:1996yc}), and have since been 
shown to arise in Chern--Simons theories interacting with parity violating matter. Parity violating (or parity odd) 
structures were studied in~\cite{Giombi:2011kc, Costa:2011mg, Aharony:2011jz, Maldacena:2012sf, 
Aharony:2012nh, GurAri:2012is, Nizami:2013tpa, Giombi:2016zwa, Aharony:2018pjn}.
Recently they were also studied in light-cone gauge \cite{Skvortsov:2018uru}, and in momentum space~\cite{Jain:2021wyn}.\footnote{Parity even correlation 
functions in momentum space were discussed in~\cite{Coriano:2013jba, Bzowski:2013sza, Bzowski:2015pba, Bzowski:2015yxv, Bzowski:2017poo, Coriano:2018bbe, 
Bautista:2019qxj, Jain:2020rmw, Jain:2020puw}.}

In contrast with the non-supersymmetric case studied in \cite{Osborn:1993cr, Erdmenger:1996yc}, supersymmetry imposes additional restrictions on the structure of three-point functions of conserved currents.
In supersymmetric field theories the energy-momentum tensor is replaced with the supercurrent multiplet \cite{Ferrara:1974pz}, which contains the 
energy-momentum tensor, the supersymmetry current and additional components such as the $R$-symmetry current.  
Similarly, a conserved vector current becomes a component of the flavour current supermultiplet. 
The general formalism to construct the two- and three-point functions of primary operators in three-dimensional 
superconformal field theories was developed in~\cite{Park:1999cw, Buchbinder:2015qsa, Buchbinder:2015wia, Kuzenko:2016cmf}.\footnote{A similar formalism
in four dimensions was developed in~\cite{Park:1997bq, Osborn:1998qu, Kuzenko:1999pi} and in six dimensions in~\cite{Park:1998nra}.} 
Within this formalism it was shown in~\cite{Buchbinder:2015qsa} that the three-point function of the supercurrent (and, hence, of the energy-momentum 
tensor) in three-dimensional $\cN=1$ superconformal theory is comprised of only one tensor structure. It was also shown that the three-point function 
of the non-abelian flavour current (and, hence, the three-point function of conserved vector currents) also contains only one tensor structure. 
In both cases the tensor structures are parity even. 

The aim of this paper is to apply the approach of~\cite{Buchbinder:2015qsa} to the case of mixed correlators
involving the supercurrent and flavour current multiplets in theories with $1 \leq \cN \leq 4$ superconformal symmetry. Our method is based on a systematic decomposition of the relevant tensors 
into irreducible components, which guarantees that all possible linearly independent structures are consistently taken into account. 
We demonstrate that these correlation functions contain only parity even structures, hence in combination with the results of \cite{Buchbinder:2015qsa} we conclude that supersymmetry forbids parity odd structures in the 
three-point functions of conserved low spin currents such as the energy momentum tensor, supersymmetry current and 
conserved vector current. In~\cite{Maldacena:2011jn} Maldacena and Zhiboedov showed under quite general assumptions 
that if a three-dimensional conformal field theory possesses a conserved higher spin current then it is free. Since a free theory results in only parity even contributions to correlation functions, we arrive at the conclusion that if the assumptions of~\cite{Maldacena:2011jn} are fulfilled, one cannot obtain 
parity odd structures in three-point functions of all conserved currents in supersymmetric conformal field theories. 

The paper is organised as follows. In section \ref{section2} we review the construction of the two-point and three-point building blocks 
which appear in correlation functions of primary superfields. We also review the general form of two- and three-point correlation functions 
of primary operators. 
In section \ref{section3} we introduce a systematic approach to solve for correlation functions of conserved currents.
We illustrate our method by reconsidering the flavour current three-point function which was previously computed in \cite{Buchbinder:2015qsa}. 
In section \ref{section4} we study three-point functions of mixed correlators involving both the supercurrent and the flavour current multiplet. 
We show that the three-point function involving one supercurrent and two flavour current multiplets is fixed by the $\cN=1$ 
superconformal symmetry up to an overall coefficient. We also show that the three-point function involving two supercurrents and 
one flavour current vanishes.
In section \ref{section5} we present a systematic discussion regarding the absence of parity violating structures in our results.
In section \ref{section6} we generalise our method to superconformal theories with $\cN=2$ supersymmetry. We show that both mixed correlators are fixed up to an overall coefficient. In section \ref{section7} we extend our analysis to the case of $\cN=3$ and $\cN=4$ superconformal symmetry. In appendix \ref{AppA} we summarise our three-dimensional notation and conventions. 

The non-vanishing of the three-point function of two supercurrents and one flavour current in $\cN=2$ theories is quite a surprise given that a similar three-point function vanishes in the $\cN=1$ case. 
Naively it appears to be a contradiction, as any theory with $\cN=2$ supersymmetry is also a theory with $\cN=1$ 
supersymmetry. From an intuitive standpoint, the number of independent tensor structures cannot grow as one increases the number of supersymmetries. 
Nevertheless, we explain that our results in the $\cN=1$ and $\cN=2$ cases are fully consistent. 


\section{Superconformal building blocks} \label{section2}


The formalism to construct correlation functions of primary operators for conformal field theories in general dimensions was 
first elucidated in \cite{Osborn:1993cr} using an efficient group theoretic formalism. In four dimensions the method was then extended to 
the case of $\cN=1$ supersymmetry in \cite{Park:1997bq,Osborn:1998qu,Park:1999pd}, and was later generalised to 
higher $\cN$ in \cite{Kuzenko:1999pi}. Here we review the pertinent details of the three-dimensional 
formalism \cite{Park:1999cw,Buchbinder:2015qsa} necessary to construct correlation functions of
 the 3D supercurrent and flavour current multiplets. 


\subsection{Superconformal transformations and primary superfields}


Let us begin by reviewing infinitesimal superconformal transformations and the transformation laws of primary superfields. This section closely follows the notation of \cite{Kuzenko:2006mv,Kuzenko:2010rp,Kuzenko:2010bd}. Consider 3D $\cN$-extended Minkowski superspace $\mathbb{M}^{3 | 2 \cN}$, parameterised by coordinates $z^{A} = (x^{a} , \q^{\a}_{I})$, where $a = 0,1,2$, $\a = 1,2$ are Lorentz and spinor indices respectively, while $I = 1,...,\cN$ is the $R$-symmetry index. 
The 3D $\cN$-extended superconformal group cannot act by smooth transformations on $\mathbb{M}^{3 | 2 \cN}$, in general only infinitesimal 
superconformal transformations are well defined. Such a transformation
\begin{equation}
	\d z^{A} = \x z^{A}  \hspace{3mm} \Longleftrightarrow \hspace{3mm} \d x^{a} = \x^{a}(z) + \text{i} (\g^{a})_{\a \b} \, \x^{\a}_{I}(z) \, \q^{\b}_{I} \, , 
	\hspace{8mm} \d \q^{\a}_{I} = \x^{\a}_{I}(z) , \, 
\label{new1}	
\end{equation}
is associated with the real first-order differential operator
\begin{equation}
	\x = \x^{A}(z) \, \partial_{A} = \x^{a}(z) \, \partial_{a} + \x^{\a}_{I}(z) D_{\a}^{I} \, , \label{Superconformal Killing vector field}
\end{equation}
which satisfies the master equation $[\x , D_{\a}^{I} ] \propto D^{J}_{\b}$. From the master equation we find
\begin{equation}
	\x^{\a}_{I} = \frac{\text{i}}{6} D_{\b I } \x^{\a \b} \, ,
\end{equation}
which implies the conformal Killing equation
\begin{equation}
	\partial_{a} \x_{b} + \partial_{b} \x_{a} = \frac{2}{3} \eta_{a b} \partial_{c} \x^{c} \, .
\label{new2}	
\end{equation}
The solutions to the master equation are called the conformal Killing supervector fields of Minkowski superspace \cite{Buchbinder:1998qv,Kuzenko:2010rp}. 
They span a Lie algebra isomorphic to the superconformal algebra $\mathfrak{osp}(\cN | 2 ; \mathbb{R})$. 
The components of the operator $\x$ were calculated explicitly in \cite{Park:1999cw}, and are found to be
\begin{subequations}
	\begin{align}
		\begin{split}
			\x^{\a \b} &= a^{\a \b} - \l^{\a}{}_{\g} x^{\g \b} - x^{\a \g} \l_{\g}{}^{\b} + \s x^{\a \b} + 4 \text{i} \e^{(\a}_{I} \q^{\b)}_{I} + 2 \text{i} \L_{IJ} \q^{\a}_{J} \q^{\b}_{I} \\
			& \hspace{10mm} + x^{\a \g} x^{\b \d} b_{\g \d} + \text{i} b_{\d}^{(\a } x^{\b) \d} \q^{2} - \frac{1}{4} b^{\a \b} \q^{2} \q^{2} - 4 \text{i} \eta_{\g I} x^{\g(\a} \q^{\b)}_{I} + 2 \eta_{I}^{(\a} \q^{\b)}_{I} \q^{2} \, , \label{Superconformal killing vector - component 1}
		\end{split}
	\end{align}
	\vspace{-5mm}
	\begin{align}	
		\x^{\a}_{I} &= \e^{\a}_{I} - \l^{\a}{}_{\b} \q^{\b}_{I} + \frac{1}{2} \s \q^{\a}_{I} + \L_{IJ} \q^{\a}_{J} + b_{\b \g} \boldsymbol{x}^{\b \a} \q^{\g}_{I} + \eta_{\b J} ( 2 \text{i} \q^{\b}_{I} \q^{\a}_{J} - \d_{IJ} \boldsymbol{x}^{\b \a} ) \, , \label{Superconformal killing vector - component 2}
	\end{align}
\end{subequations}
\begin{equation}
	a_{\a \b} = a_{\b \a} \, , \hspace{5mm} \l_{\a \b} = \l_{\b \a} \, , \hspace{2mm} \l^{\a}{}_{\a} = 0 \, , \hspace{5mm} b_{\a \b} = b_{\b \a} \, , \hspace{5mm} \L_{IJ}= - \L_{JI} \, .
\end{equation}
The bosonic parameters $a_{\a \b}$, $\l_{\a \b}$, $\s$, $b_{\a \b}$, $\L_{IJ}$ correspond to infinitesimal translations, 
Lorentz transformations, scale transformations, special conformal transformations and $R$-symmetry transformations respectively, while the
fermionic parameters $\e^{\a}_{I}$ and $\eta^{\a}_{I}$ correspond to $Q$-supersymmetry and $S$-supersymmetry transformations. 
Furthermore, the identities
\begin{equation}
	D^{I}_{[\a} \x^{J}_{\b]} \propto \ve_{\a \b} \, , \hspace{8mm} D^{I}_{(\a} \x^{J}_{\b)} \propto \d^{IJ} \, , \hspace{8mm} D^{(I}_{[\a} \x^{J)}_{\b]} \propto \d^{IJ} \ve_{\a \b} \, ,
	\label{new3}
\end{equation}
imply that 
\begin{equation}
	[\x , D_{\a}^{I} ] = - ( D^{I}_{\a} \x^{\b}_{J}) D^{J}_{\b} = \l_{\a}{}^{\b}(z) D^{I}_{\b} + \L^{IJ}(z) D^{J}_{\a} - \frac{1}{2} \s(z) D^{I}_{\a} \, ,
\end{equation}
\begin{equation}
	\l_{\a \b}(z) = - \frac{1}{\cN} D^{I}_{(\a} \x^{I}_{\b)} \, , \hspace{5mm} \L^{IJ}(z) = - 2 D^{[I}_{\a} \x^{J] \a} \, , \hspace{5mm} \s(z) = \frac{1}{\cN} D^{I}_{\a} \x^{\a}_{I} \, . 
	\label{new4}
\end{equation}
The local parameters $\l^{\a \b}(z)$, $\L_{IJ}(z)$, $\s(z)$ are interpreted as being associated with combined special-conformal/Lorentz, $R$-symmetry and scale transformations respectively, and appear in the transformation laws for primary tensor superfields. For later use let's also introduce the $z$-dependent $S$-supersymmetry parameter
\begin{equation}
	\eta_{I \a}(z) = -\frac{\text{i}}{2} D_{I \a} \s(z) \,.
	\label{new5}
\end{equation}
Explicit calculations of the local parameters give \cite{Park:1999cw}
\begin{subequations}
	\begin{align}
		\l^{\a \b}(z) &= \l^{\a \b} - x^{\g (\a} b^{\b)}_{\g} - \frac{\text{i}}{2} b^{\a \b} \q^{2} + 2 \text{i} \eta^{(\a}_{I} \q^{\b)}_{I} \, , \label{Local parameter 1} \\ 
		\L_{IJ}(z) &= \L_{IJ} + 4 \text{i} \eta^{\a}_{[I} \q_{J] \a} + 2 \text{i} b_{\a \b} \q^{\a}_{I} \q^{\b}_{J} \, , \label{Local parameter 2} \\[2mm]
		\s(z) &= \s + b_{\a \b} x^{\a \b} + 2 \text{i} \q^{\a}_{I} \eta_{\a I} \, , \label{Local parameter 3} \\[2mm]
		\eta_{\a I}(z) &= \eta_{\a I} - b_{\a \b} \q^{\b}_{I} \, . \label{Local parameter 4}
	\end{align}
\end{subequations}
Now consider a generic tensor superfield $\F^{\cI}_{\cA}(z)$ transforming in a representation $T$ of the Lorentz group with respect to the index $\cA$, 
and in the representation $D$ of the $R$-symmetry group $\sO(\cN)$ with respect to the index 
$\cI$.\footnote{We assume the representations $T$ and $D$ are irreducible.} Such a superfield is called primary with dimension $q$ if its superconformal transformation law is 
\begin{equation}
	\d \F^{\cI}_{\cA} = - \x \F^{\cI}_{\cA} - q \s(z) \F^{\cI}_{\cA} + \l^{\a \b}(z) (M_{\a \b})_{\cA}{}^{\cB} \F^{\cI}_{\cB} + 
	\L^{I J}(z) (R_{I J})^{\cI}{}_{\cJ} \F^{\cJ}_{\cA} \,,
	\label{new6}
\end{equation}
where $\x$ is the superconformal Killing vector, $\s(z)$, $\l^{\a \b}(z)$, $\L_{I J}(z)$ are the $z$-dependent parameters associated with $\x$, and the matrices $M_{\a \b}$ and $R_{IJ}$ are the Lorentz and $\sO(\cN)$ generators respectively.


\subsection{Two-point functions}


Given two superspace points $z_{1}$ and $z_{2}$, we can define the two-point functions
\begin{equation}
	\boldsymbol{x}_{12}^{\alpha \beta} = (x_{1} - x_{2})^{\alpha \beta} + 2 \text{i} \theta^{(\alpha}_{1 I} \theta^{\beta)}_{2 I} - \text{i} \theta^{\a}_{12 I} \theta^{\b}_{12 I} \, ,  \hspace{10mm} \theta^{\alpha I}_{12} = \theta_{1}^{\alpha I} - \theta_{2}^{\alpha I} \, , \label{Two-point building blocks 1}
\end{equation}
which transform under the superconformal group as follows
\begin{subequations}
	\begin{align}
		\tilde{\d} \boldsymbol{x}_{12}^{\a \b} &= \bigg( \frac{1}{2} \d^{\a}{}_{\g} \, \s(z_{1}) - \l^{\a}{}_{\g}(z_{1}) \bigg) \boldsymbol{x}_{12}^{\g \b} + \boldsymbol{x}_{12}^{\a \g} \bigg( \frac{1}{2} \d_{\g}{}^{\b} \s(z_{2}) - \l_{\g}{}^{\b}(z_{2}) \bigg) \, , \label{Two-point building blocks 1 - transformation law 1} \\[2mm]
		\tilde{\d} \q_{12 \, I}^{\a} &= \bigg( \frac{1}{2} \d^{\a}{}_{\b} \, \s(z_{1}) - \l^{\a}{}_{\b}(z_{1}) \bigg) \q_{12 \, I}^{\b} - \boldsymbol{x}_{12}^{\a \b} \, \eta_{\b I}(z_{2}) + \L_{IJ}(z_{2}) \, \q_{12 \, J}^{\a} \,. \label{Two-point building blocks 1 - transformation law 2}
	\end{align}
\end{subequations}
Here the total variation $\tilde{\d}$ is defined by its action on an $n$-point function $\F(z_{1},...,z_{n})$ as
\begin{equation}
	\tilde{\d} \F(z_{1},...,z_{n}) = \sum_{i=1}^{n} \x_{z_{i}} \F(z_{1},...,z_{n}) \, . \label{Total variation} 
\end{equation}
It should be noted that~\eqref{Two-point building blocks 1 - transformation law 2} 
contains an inhomogeneous piece in its transformation law, hence it will not appear as a building block in two- or three-point functions. 
Due to the useful property, $\boldsymbol{x}_{21}^{\a \b} = - \boldsymbol{x}_{12}^{\b \a}$, the two-point function \eqref{Two-point building blocks 1} can be split into symmetric and antisymmetric parts as follows
\begin{equation}
	\boldsymbol{x}_{12}^{\a \b} = x_{12}^{\a \b} + \frac{\text{i}}{2} \ve^{\alpha \beta} \theta^{2}_{12} \, , \hspace{10mm} \q_{12}^{2} = \q_{12 I}^{\a} \q_{12 \a I} \, . \label{Two-point building blocks 1 - properties 1}
\end{equation}
The symmetric component
\begin{equation}
	x_{12}^{\a \b} = (x_{1} - x_{2})^{\alpha \beta} + 2 \text{i} \theta^{(\alpha}_{1 I} \theta^{\beta)}_{2 I} \, , \label{Two-point building blocks 1 - properties 2}
\end{equation}
is recognised as the bosonic part of the standard two-point superspace interval. Next let us introduce the two-point objects
\begin{subequations}
	\begin{align}
		\boldsymbol{x}_{12}^{2} &= -\frac{1}{2} \boldsymbol{x}_{12}^{\a \b} \boldsymbol{x}_{12 \a \b} \, , \hspace{5mm} \label{Two-point building blocks 2} \\
		\hat{\boldsymbol{x}}_{12}^{\a \b} = \frac{\boldsymbol{x}_{12}^{\a \b}}{\sqrt{ \boldsymbol{x}_{12}^{2}}} &\, , \hspace{10mm} \hat{\boldsymbol{x}}_{12 \a}{}^{\g} \hat{\boldsymbol{x}}_{12 \g}{}^{\b} = \d_{\a}{}^{\b} \, . \label{Two-point building blocks 3}
	\end{align}
\end{subequations}
Hence, we find
\begin{equation}
	(\boldsymbol{x}_{12}^{-1})^{\a \b} = - \frac{\boldsymbol{x}_{12}^{\b \a}}{\boldsymbol{x}_{12}^{2}} \, . \label{Two-point building blocks 4}
\end{equation}
Under superconformal transformations, \eqref{Two-point building blocks 2} transforms with local scale parameters, while \eqref{Two-point building blocks 3} transforms with local Lorentz parameters
\begin{subequations}
	\begin{align}
		\tilde{\d} \boldsymbol{x}_{12}^{2} &= ( \s(z_{1}) + \s(z_{2}) ) \, \boldsymbol{x}_{12}^{2} \, , \label{Two-point building blocks 2 - transformation law 1} \\
		\tilde{\d} \hat{\boldsymbol{x}}_{12}^{\a \b} &= - \l^{\a}{}_{\g}(z_{1}) \, \hat{\boldsymbol{x}}_{12}^{\g \b} - \hat{\boldsymbol{x}}_{12}^{\a \g} \, \l_{\g}{}^{\b}(z_{2}) \, . \label{Two-point building blocks 3 - transformation law 1}
	\end{align}
\end{subequations}
Thus, both objects are essential in the construction of correlation functions of primary superfields. We also have the useful differential identities
\begin{equation}
	D^{I}_{(1) \g} \boldsymbol{x}_{12}^{\a \b} = - 2 \text{i} \q^{I \b}_{12} \d_{\g}^{\a} \, , \hspace{10mm} D^{I}_{(1) \a} \boldsymbol{x}_{12}^{\a \b} = - 4 \text{i} \q^{I \b}_{12} \, , \label{Two-point building blocks 1 - differential identities}
\end{equation}
where $D^{I}_{(i) \a}$ is the standard covariant spinor derivative \eqref{Covariant spinor derivatives} acting on the superspace point $z_{i}$. 
Finally, for completeness, the $\text{SO}(\cN)$ structure of primary superfields in correlation functions is addressed by the $\cN \times \cN$ matrix
\begin{equation}
	u_{12}^{IJ} = \d^{IJ} + 2 \text{i} \q_{12}^{I \a} (\boldsymbol{x}_{12}^{-1})_{\a \b} \q_{12}^{J \b} \, , \label{Two-point building blocks 5}
\end{equation}
which is orthogonal and unimodular,
\begin{equation}
	u_{12}^{IK} u_{12}^{KJ} = \d^{IJ} \, , \hspace{10mm} \det u_{12} = 1 \, . \label{Two-point building blocks 5 - properties}
\end{equation}
The infinitesimal variation of this matrix is 
\begin{equation}
	\tilde{\d} u_{12}^{IJ} = \L^{IK}(z_{1}) \, u_{12}^{KJ} - u_{12}^{IK} \L^{KJ}(z_{2}) \, . \label{Two-point building blocks 5 - transformation law 1}
\end{equation}
Hence,~\eqref{Two-point building blocks 5} is expected to appear in the construction of correlation functions of primary superfields 
with $\text{SO}(\cN)$ indices. 

The two-point correlation function of a primary superfield $\F^{\cI}_{\cA}$ and its conjugate $\bar{\F}^{\cB}_{\cJ}$ is fixed by the superconformal symmetry 
as follows
\begin{equation}
	\langle \F^{\cI}_{\cA}(z_{1}) \bar{\F}^{\cB}_{\cJ}(z_{2}) \rangle = c \, \frac{T_{\cA}{}^{\cB}(\hat{\boldsymbol{x}}_{12}) D^{\cI}{}_{\cJ}(u_{12})}{(\boldsymbol{x}_{12}^{2})^{q}} \, , 
\end{equation} 
where $c$ is a constant coefficient. The denominator of the two-point function is determined by the conformal dimension of $\F^{\cI}_{\cA}$, 
which guarantees that the correlation function transforms with the appropriate weight under scale transformations.


\subsection{Three-point functions}


Given three superspace points $z_{i}$, $i = 1,2,3$, one can define the three-point building blocks $\cZ_{i} = ( \boldsymbol{X}_{i} , \Q_{i} )$ as follows:
\begin{subequations} \label{Three-point building blocks 1}
\begin{align}
		\boldsymbol{X}_{1 \, \a \b} &= -(\boldsymbol{x}_{21}^{-1})_{\a \g}  \boldsymbol{x}_{23}^{\g \d} (\boldsymbol{x}_{13}^{-1})_{\d \b} \, , \hspace{5mm} \Q^{I}_{1 \, \a} = (\boldsymbol{x}_{21}^{-1})_{\a \b} \q_{12}^{I \b} - (\boldsymbol{x}_{31}^{-1})_{\a \b} \q_{13}^{I \b} \, , \label{Three-point building blocks 1a} \\[2mm]
		\boldsymbol{X}_{2 \, \a \b} &= -(\boldsymbol{x}_{32}^{-1})_{\a \g}  \boldsymbol{x}_{31}^{\g \d} (\boldsymbol{x}_{21}^{-1})_{\d \b} \, , \hspace{5mm} \Q^{I}_{2 \, \a} = (\boldsymbol{x}_{32}^{-1})_{\a \b} \q_{23}^{I \b} - (\boldsymbol{x}_{12}^{-1})_{\a \b} \q_{21}^{I \b} \, , \label{Three-point building blocks 1b} \\[2mm]
		\boldsymbol{X}_{3 \, \a \b} &= -(\boldsymbol{x}_{13}^{-1})_{\a \g}  \boldsymbol{x}_{12}^{\g \d} (\boldsymbol{x}_{32}^{-1})_{\d \b} \, , \hspace{5mm} \Q^{I}_{3 \, \a} = (\boldsymbol{x}_{13}^{-1})_{\a \b} \q_{31}^{I \b} - (\boldsymbol{x}_{23}^{-1})_{\a \b} \q_{32}^{I \b} \, . \label{Three-point building blocks 1c}
\end{align}
\end{subequations}
These objects, along with their corresponding transformation laws, may be obtained from one-another by cyclic permutation of superspace points. 
The building blocks transform covariantly under the action of the superconformal group:
\begin{subequations}
	\begin{align}
		\tilde{\d} \boldsymbol{X}_{1 \, \a \b} &= \l_{\a}{}^{\g}(z_{1}) \boldsymbol{X}_{1 \, \g \b} + \boldsymbol{X}_{1 \, \a \g} \l^{\g}{}_{\b}(z_{1}) - \s(z_{1}) \boldsymbol{X}_{1 \, \a \b} \, , \label{Three-point building blocks 1a - transformation law 1} \\[2mm]
		\tilde{\d} \Q^{I}_{1 \, \a} &= \bigg( \l_{\a}{}^{\b}(z_{1}) - \frac{1}{2} \d_{\a}{}^{\b} \s(z_{1}) \bigg) \Q^{I}_{1 \, \b} + \L^{IJ}(z_{1}) \, \Q^{J}_{1 \, \a} \, . \label{Three-point building blocks 1a - transformation law 2}
	\end{align}
\end{subequations}
Therefore~\eqref{Three-point building blocks 1a},~\eqref{Three-point building blocks 1b} and~\eqref{Three-point building blocks 1c} will appear 
as building blocks in three-point correlations functions. It should be noted that under scale transformations of superspace, 
$z^{A} = (x^{a},\q^{\a}) \mapsto z'^{A} = (\l^{-2} x^{a} , \l^{-1} \q^{\a})$, the three-point building blocks transform as 
$\cZ = (\boldsymbol{X},\Q) \mapsto \cZ' = (\l^{2} \boldsymbol{X}, \l \Q)$. 
Next we define
\begin{equation}
	\boldsymbol{X}_{1}^{2} = - \frac{1}{2} \boldsymbol{X}_{1}^{\a \b}  \boldsymbol{X}_{1 \, \a \b} = \frac{\boldsymbol{x}_{23}^{2}}{\boldsymbol{x}_{13}^{2} \boldsymbol{x}_{12}^{2}} \, , \hspace{10mm}  \Q_{1}^{2} = \Q^{I \a}_{1} \Q^{I}_{1 \, \a} \, , \label{Three-point building blocks 2}
\end{equation}
which, due to~\eqref{Three-point building blocks 1a - transformation law 1} and~\eqref{Three-point building blocks 1a - transformation law 2}, have the transformation laws
\begin{equation}
	\tilde{\d} \boldsymbol{X}_{1}^{2} = - 2 \s(z_{1}) \boldsymbol{X}_{1}^{2} \, , \hspace{10mm} \tilde{\d} \Q_{1}^{2} = - \s(z_{1}) \, \Q_{1}^{2} \, . \label{Three-point building blocks 2 - transformation law 1}
\end{equation}
We also define the inverse of $\boldsymbol{X}_{1}$,
\begin{equation}
	(\boldsymbol{X}_{1}^{-1})^{\a \b} = - \frac{\boldsymbol{X}_{1}^{\b \a}}{\boldsymbol{X}_{1}^{2}} \, ,
\end{equation}  
and introduce useful identities involving $\boldsymbol{X}_{i}$ and $\Q_{i}$ at different superspace points, e.g.,
\begin{subequations}
	\begin{align}
		\boldsymbol{x}_{13}^{\a \a'} \boldsymbol{X}_{3 \, \a' \b'} \boldsymbol{x}_{31}^{\b' \b} &= - (\boldsymbol{X}_{1}^{-1})^{\b \a} \, , \label{Three-point building blocks 1a - properties 1}\\[2mm]
		\Q_{1 \, \g}^{I} \boldsymbol{x}_{13}^{\g \d} \boldsymbol{X}_{3 \, \d \b} &= u_{13}^{IJ} \Q_{3 \, \b}^{J} \, . \label{Three-point building blocks 1a - properties 2}
	\end{align}
\end{subequations}
As a consequence of \eqref{Three-point building blocks 2 - transformation law 1}, we can identify the three-point superconformal invariant
\begin{equation}
	\frac{\Q_{1}^{2}}{\sqrt{\boldsymbol{X}_{1}^{2}}} \hspace{5mm} \Rightarrow \hspace{5mm} \tilde{\d} \bigg( \frac{\Q_{1}^{2}}{\sqrt{\boldsymbol{X}_{1}^{2}}} \bigg) = 0 \, .
\end{equation}
Hence, the superconformal symmetry fixes the functional form of three-point correlation functions up to this combination. Indeed, 
using~\eqref{Three-point building blocks 1a - properties 1} and~\eqref{Three-point building blocks 1a - properties 2} one can show that the superconformal invariant is also invariant under permutation of superspace points, i.e 
\begin{equation}
	\frac{\Q_{1}^{2}}{\sqrt{\boldsymbol{X}_{1}^{2}}} = \frac{\Q_{2}^{2}}{\sqrt{\boldsymbol{X}_{2}^{2}}} = \frac{\Q_{3}^{2}}{\sqrt{\boldsymbol{X}_{3}^{2}}}  \, . \label{Superconformal invariants}
\end{equation}
The three-point objects~\eqref{Three-point building blocks 1a},~\eqref{Three-point building blocks 1b} and~\eqref{Three-point building blocks 1c} have many properties similar to those of the two-point building blocks. After decomposing $\boldsymbol{X}_1$ 
into symmetric and antisymmetric parts similar to \eqref{Two-point building blocks 1 - properties 1} we have
\begin{equation}
	\boldsymbol{X}_{1 \, \a \b} = X_{1 \, \a \b} - \frac{\text{i}}{2} \ve_{\a \b} \Q_{1}^{2} \, , \hspace{10mm} X_{1 \, \a \b} = X_{1 \, \b \a} \, , \label{Three-point building blocks 1a - properties 3}
\end{equation}
where the symmetric spinor $X_{1 \, \a \b}$ can be equivalently represented by the three-vector $X_{1 \, m} = - \frac{1}{2} (\g_{m})^{\a \b} X_{1 \, \a \b}$. It is now convenient to introduce analogues of the covariant spinor derivative and supercharge operators involving the three-point objects,
\begin{equation}
	\cD^{I}_{(1) \a} = \frac{\partial}{\partial \Q^{\a}_{1 I}} + \text{i} (\g^{m})_{\a \b} \Q^{I \b}_{1} \frac{\partial}{\partial X^{m}_{1}} \, , \hspace{5mm} \cQ^{I}_{(1) \a} = \text{i} \frac{\partial}{\partial \Q^{\a}_{1 I}} + (\g^{m})_{\a \b} \Q^{I \b}_{1} \frac{\partial}{\partial X^{m}_{1}} \, , \label{Supercharge and spinor derivative analogues}
\end{equation}
which obey the standard commutation relations
\begin{equation}
	\big\{ \cD^{I}_{(i) \a} , \cD^{J}_{(i) \b} \big\} = \big\{ \cQ^{I}_{(i) \a} , \cQ^{J}_{(i) \b} \big\} = 2 \text{i} \, \d^{IJ} (\g^{m})_{\a \b} \frac{\partial}{\partial X^{m}_{i}} \, .
\end{equation}
Some useful identities involving~\eqref{Supercharge and spinor derivative analogues} are
\begin{equation}
	\cD^{I}_{(1) \g} \boldsymbol{X}_{1 \, \a \b} = - 2 \text{i} \ve_{\g \b} \Q^{I}_{1 \, \a} \, , \hspace{5mm} \cQ^{I}_{(1) \g} \boldsymbol{X}_{1 \, \a \b} = - 2 \ve_{\g \a} \Q^{I}_{1 \, \b} \, . \label{Three-point building blocks 1a - differential identities 1}
\end{equation}
We must also account for the fact that various primary superfields obey certain differential equations. Using \eqref{Two-point building blocks 1 - differential identities} we arrive at the following
\begin{subequations}
	\begin{align}
		D_{(1) \g}^{I} \boldsymbol{X}_{3 \, \a \b} &= 2 \text{i} (\boldsymbol{x}^{-1}_{13})_{\a \g} u_{13}^{IJ} \Q_{3 \, \b}^{J} \, , \hspace{5mm} D_{(1) \a}^{I} \Q_{3 \, \b}^{J} = - (\boldsymbol{x}_{13}^{-1})_{\b \a} u_{13}^{IJ} \, , \label{Three-point building blocks 1c - differential identities 1}\\[2mm]
		D_{(2) \g}^{I} \boldsymbol{X}_{3 \, \a \b} &= 2 \text{i} (\boldsymbol{x}^{-1}_{23})_{\b \g} u_{23}^{IJ} \Q_{3 \, \b}^{J} \, , \hspace{5mm} D_{(2) \a}^{I} \Q_{3 \, \b}^{J} = (\boldsymbol{x}_{23}^{-1})_{\b \a} u_{23}^{IJ} \, . \label{Three-point building blocks 1c - differential identities 2}
	\end{align}
\end{subequations}
Now given a function $f(\boldsymbol{X}_{3} , \Q_{3})$, there are the following differential identities which arise as a consequence of \eqref{Three-point building blocks 1a - differential identities 1}, \eqref{Three-point building blocks 1c - differential identities 1} and \eqref{Three-point building blocks 1c - differential identities 2}:
\begin{subequations}
	\begin{align}
		D^{I}_{(1) \g} f(\boldsymbol{X}_{3} , \Q_{3}) &= (\boldsymbol{x}_{13}^{-1})_{\a \g} u_{13}^{IJ} \cD_{(3)}^{J \a} f(\boldsymbol{X}_{3} , \Q_{3}) \, ,  \label{Three-point building blocks 1c - differential identities 3} \\[2mm]
		D^{I}_{(2) \g} f(\boldsymbol{X}_{3} , \Q_{3}) &= \text{i} (\boldsymbol{x}_{23}^{-1})_{\a \g} u_{23}^{IJ} \cQ_{(3)}^{J \a} f(\boldsymbol{X}_{3} , \Q_{3}) \, .  \label{Three-point building blocks 1c - differential identities 4}
	\end{align}
\end{subequations}
These will prove to be essential for imposing differential constraints on correlation functions, e.g. those arising from conservation equations in the case of correlators involving the supercurrent and flavour current multiplets.

Finally, for completeness, let us introduce the three-point objects which take care of the $R$-symmetry structure of correlation functions. We define 
\begin{equation}
	U_{1}^{IJ} = u_{12}^{IK} u_{23}^{KL} u_{31}^{LJ} = \d^{IJ} + 2 \text{i} \Q_{1 \, \a}^{I} (\boldsymbol{X}_{1}^{-1})^{\a \b} \Q_{1 \, \b}^{J} \, , \label{Three-point building blocks 3}
\end{equation}
which transforms as an $\text{O}(\cN)$ tensor at $z_{1}$,
\begin{equation}
	\tilde{\d} U_{1}^{IJ} = \L^{IK}(z_{1}) \, U_{1}^{KJ} - U_{1}^{IK} \L^{KJ}(z_{1}) \, .
\end{equation}
and is orthogonal and unimodular by construction. The others are obtained by cyclic permutation of superspace points, and are related by the useful identities
\begin{equation}
	U_{2}^{IJ} = u_{21}^{IK} U_{1}^{KL} u_{12}^{LJ} \, , \hspace{10mm} U_{3}^{IJ} = u_{31}^{IK} U_{1}^{KL} u_{13}^{LJ} \, .
\end{equation}

As concerns three-point correlation functions; let $\F$, $\J$, $\P$ be primary superfields with conformal dimensions $q_{1}$, $q_{2}$ and $q_{3}$ respectively. The three-point function may be constructed using the general expression
\begin{align}
	\langle \F^{\cI_{1}}_{\cA_{1}}(z_{1}) \, \J^{\cI_{2}}_{\cA_{2}}(z_{2}) \, \P^{\cI_{3}}_{\cA_{3}}(z_{3}) \rangle =& \label{Three-point function - general ansatz} \\[2mm]
	& \hspace{-35mm} \frac{ T^{(1)}{}_{\cA_{1}}{}^{\cB_{1}}(\hat{\boldsymbol{x}}_{13}) T^{(2)}{}_{\cA_{2}}{}^{\cB_{2}}(\hat{\boldsymbol{x}}_{23}) D^{(1) \, \cI_{1}}{}_{\cJ_{1}}(u_{13}) D^{(2) \, \cI_{2}}{}_{\cJ_{2}}( u_{23}) }{(\boldsymbol{x}_{13}^{2})^{q_{1}} (\boldsymbol{x}_{23}^{2})^{q_{2}}}
	\; \cH^{\cJ_{1} \cJ_{2} \cI_{3}}_{\cB_{1} \cB_{2} \cA_{3}}(\boldsymbol{X}_{3}, \Q_{3}, U_{3}) \, , \nonumber
\end{align} 
where the tensor $\cH^{\cI_{1} \cI_{2} \cI_{3}}_{\cA_{1} \cA_{2} \cA_{3}}$ is highly constrained by the superconformal symmetry as follows:
\begin{enumerate}
\item[\textbf{(i)}] Under scale transformations of superspace the correlation function transforms as 
\begin{equation}
	\langle \F^{\cI_{1}}_{\cA_{1}}(z_{1}') \, \J^{\cI_{2}}_{\cA_{2}}(z_{2}') \, \P^{\cI_{3}}_{\cA_{3}}(z_{3}') \rangle = (\l^{2})^{q_{1} + q_{2} + q_{3}} \langle \F^{\cI_{1}}_{\cA_{1}}(z_{1}) \, \J^{\cI_{2}}_{\cA_{2}}(z_{2}) \,  \P^{\cI_{3}}_{\cA_{3}}(z_{3}) \rangle \, ,
\end{equation}
which implies that $\cH$ obeys the scaling property
\begin{equation}
	\cH^{\cI_{1} \cI_{2} \cI_{3}}_{\cA_{1} \cA_{2} \cA_{3}}(\l^{2} \boldsymbol{X}, \l \Q, U) = (\l^{2})^{q_{3} - q_{2} - q_{1}} \, \cH^{\cI_{1} \cI_{2} \cI_{3}}_{\cA_{1} \cA_{2} \cA_{3}}(\boldsymbol{X}, \Q, U) \, , \hspace{5mm} \forall \l \in \mathbb{R} \, \backslash \, \{ 0 \} \, .
\end{equation}
This guarantees that the correlation function transforms correctly under conformal transformations.
	
\item[\textbf{(ii)}] If any of the fields $\F$, $\J$, $\P$ obey differential equations, such as conservation laws in the case of conserved current multiplets, then the tensor $\cH$ is also constrained by differential equations. Such constraints may be derived with the aid of identities \eqref{Three-point building blocks 1c - differential identities 3}, \eqref{Three-point building blocks 1c - differential identities 4}.
	
\item[\textbf{(iii)}] If any (or all) of the superfields $\F$, $\J$, $\P$ coincide, the correlation function possesses symmetries under permutations of superspace points, e.g.
	\begin{equation}
		\langle \F^{\cI_{1}}_{\cA_{1}}(z_{1}) \, \F^{\cI_{2}}_{\cA_{2}}(z_{2}) \, \P^{\cI_{3}}_{\cA_{3}}(z_{3}) \rangle = (-1)^{\e(\F)} \langle \F^{\cI_{2}}_{\cA_{2}}(z_{2}) \, \F^{\cI_{1}}_{\cA_{1}}(z_{1}) \, \P^{\cI_{3}}_{\cA_{3}}(z_{3}) \rangle \, ,
\end{equation}
where $\e(\F)$ is the Grassmann parity of $\F$. As a consequence, the tensor $\cH$ obeys constraints which will be referred to as ``point-switch identities". To analyse these constraints, we note that under permutations of any two superspace points, the three-point building blocks transform as
\begin{subequations}
		\begin{align}
			\boldsymbol{X}_{3 \, \a \b} &\stackrel{1 \leftrightarrow 2}{\longrightarrow} - \boldsymbol{X}_{3 \, \b \a} \, , \hspace{10mm} \Q^{I}_{3 \, \a} \stackrel{1 \leftrightarrow 2}{\longrightarrow} - \Q^{I}_{3 \, \a} \, , \label{Three-point building blocks 1c - properties 1} \\[2mm]
			\boldsymbol{X}_{3 \, \a \b} &\stackrel{2 \leftrightarrow 3}{\longrightarrow} - \boldsymbol{X}_{2 \, \b \a} \, , \hspace{10mm} \Q^{I}_{3 \, \a} \stackrel{2 \leftrightarrow 3}{\longrightarrow} - \Q^{I}_{2 \, \a} \, , \label{Three-point building blocks 1c - properties 2} \\[2mm]
			\boldsymbol{X}_{3 \, \a \b} &\stackrel{1 \leftrightarrow 3}{\longrightarrow} - \boldsymbol{X}_{1 \, \b \a} \, , \hspace{10mm} \Q^{I}_{3 \, \a} \stackrel{1 \leftrightarrow 3}{\longrightarrow} - \Q^{I}_{1 \, \a} \, . \label{Three-point building blocks 1c - properties 3}
		\end{align}
	\end{subequations}
\end{enumerate}

The constraints above fix the functional form of $\cH$ (and therefore the correlation function) up to finitely many parameters. Hence the procedure described above reduces the problem of computing three-point correlation functions to deriving the tensor $\cH$ subject to the above constraints. In the next sections, we will apply this formalism to compute three-point correlation functions involving the supercurrent and flavour current multiplets.

\newpage


\section{Correlation functions of conserved currents in \texorpdfstring{$\cN=1$}{N=1} superconformal field theory}\label{section3}


\subsection{Supercurrent and flavour current multiplets}


The 3D, $\cN=1$ conformal supercurrent is a primary, dimension $5/2$ totally symmetric spin-tensor $J_{\a \b \g}$, which contains the three-dimensional 
energy-momentum tensor along with the supersymmetry current \cite{Buchbinder:1998qv,Komargodski:2010rb,Korovin:2016tsq}. It obeys the conservation equation
\begin{equation}
	D^{\a} J_{\a \b \g} = 0 \, ,
\end{equation}
and has the following superconformal transformation law:
\begin{equation}
	\d J_{\a \b \g} = - \x J_{\a \b \g} - \frac{5}{2} \s(z) J_{\a \b \g} + 3 \l(z)^{\d}_{(\a} J_{\b \g) \d} \, .
\end{equation}
The $\cN=1$ supercurrent may be derived from, for example, supergravity prepotential approaches \cite{Buchbinder:1998qv} or the superfield Noether procedure \cite{Magro:2001aj,Kuzenko:2010ni}. 

The general formalism in section \ref{section2} allows the two-point function to be determined up to a single real coefficient:
\begin{equation}
	\langle J_{\a \b \g}(z_{1}) J^{\a' \b' \g'}(z_{2}) \rangle = \text{i} b_{\cN=1} \frac{\boldsymbol{x}_{12 (\a}{}^{\a'} \boldsymbol{x}_{12 \b}{}^{\b'} \boldsymbol{x}_{12 \g)}{}^{\g'}}{(\boldsymbol{x}_{12}^{2})^{4}} \, .
\end{equation}
It is then a simple exercise to show that the two-point function has the right symmetry properties under permutation of superspace points
\begin{equation}
	\langle J_{\a \b \g}(z_{1}) J_{\a' \b' \g'}(z_{2}) \rangle = - \langle J_{\a' \b' \g'}(z_{2}) J_{\a \b \g}(z_{1}) \rangle \, ,
\end{equation}
and also satisfies
\begin{equation}
	D_{(1)}^{\a} \langle J_{\a \b \g}(z_{1}) J_{\a' \b' \g'}(z_{2}) \rangle = 0 \, .
\end{equation}
Next let's consider the 3D $\cN=1$ flavour current, which is represented by a primary, dimension $3/2$ spinor superfield 
$L_{\a}$ obeying the conservation equation\footnote{The tensor structure 
and the conservation law of the $1 \leq\cN \leq 4$ flavour currents  
follow from the structure of unconstrained prepotentials for $1 \leq\cN \leq 4$ 
vector multiplets~\cite{Siegel:1979fr, Gates:1983nr, Hitchin:1986ea, Zupnik:1988en, Zupnik:1988wa, Zupnik:1999iy, Zupnik:2009zn}.}
\begin{equation}
	D^{\a} L_{\a} = 0 \, . \label{N=1 flavour current conservation equation}
\end{equation}
It transforms covariantly under the superconformal group as
\begin{equation}
	\d L_{\a} = -\x L_{\a} - \frac{3}{2} \s(z) L_{\a} + \l(z)_{\a}{}^{\b} L_{\b} \, . 
\end{equation}
We can also consider the case when there are several flavour current multiplets (represented by the flavour index, $\abar$) corresponding to a simple flavour group. According to general formalism in section 2, the two-point function for $\cN=1$ flavour current multiplets is fixed up to a single real coefficient $a_{\cN=1}$
\begin{equation}
	\langle L^{\abar}_{\a}(z_{1}) L^{\bbar}_{\b}(z_{2}) \rangle = \text{i} a_{\cN=1} \frac{\d^{\abar \bbar} \boldsymbol{x}_{12 \a \b} }{(\boldsymbol{x}_{12}^{2})^{2}} \, .
\end{equation}
It is easy to see that the two-point function obeys the correct symmetry properties under permutation of superspace points, $\langle L^{\bar{a}}_{\a}(z_{1}) L^{\bar{b}}_{\b}(z_{2}) \rangle = - \langle L^{\bar{b}}_{\b}(z_{2}) L^{\bar{a}}_{\a}(z_{1}) \rangle$. 
One can also check that it satisfies the conservation equation~\eqref{N=1 flavour current conservation equation}
\begin{equation}
	D_{(1)}^{\a} \langle L^{\bar{a}}_{\a}(z_{1}) L^{\bar{b}}_{\b}(z_{2}) \rangle = 0 \, .
\end{equation}
Three-point correlation functions of the flavour current and particularly the supercurrent are considerably more complicated, and were derived in \cite{Buchbinder:2015qsa,Buchbinder:2015wia}. However, correlators of \textit{combinations} of these fields (mixed correlators) were not studied previously and will be analysed in section \ref{section4}.


\subsection{Correlation functions of conserved current multiplets}


The possible three-point correlation functions that may be constructed from the conserved $\cN=1$ supercurrent and flavour current multiplets are:
\begin{align}
	&\langle L^{\abar}_{\a}(z_{1}) L^{\bbar}_{\b}(z_{2}) L^{\cbar}_{\g}(z_{3}) \rangle \, , \hspace{10mm} \langle J_{\cA}(z_{1}) J_{\cB}(z_{2}) J_{\cC}(z_{3}) \rangle \, , \\
	&\langle L^{\abar}_{\a}(z_{1}) J_{\cA}(z_{2}) L^{\bbar}_{\b}(z_{3}) \rangle \, , \hspace{10mm} \langle J_{\cA}(z_{1}) J_{\cB}(z_{2}) L^{\abar}_{\a}(z_{3})  \rangle \, ,
\end{align}
where $\cA, \cB, \cC$ each denote a totally symmetric combination of three spinor indices. 
The correlators $\langle L^{\abar}_{\a}(z_{1}) L^{\bbar}_{\b}(z_{2}) L^{\cbar}_{\g}(z_{3}) \rangle$ and 
$\langle J_{\cA}(z_{1}) J_{\cB}(z_{2}) J_{\cC}(z_{3}) \rangle$ were studied in \cite{Buchbinder:2015qsa}. 
Before we compute the mixed correlators, let us demonstrate our method on 
the three-point function $\langle L^{\abar}_{\a}(z_{1}) L^{\bbar}_{\b}(z_{2}) L^{\cbar}_{\g}(z_{3}) \rangle$, which is comparatively straightforward.

The general form of the flavour current three-point function is:\footnote{Here we consider only the contribution proportional 
to the totally antisymmetric structure constants $f^{\abar \bbar \cbar}$. Similarly, one can consider the  contribution 
totally symmetric in flavour indices. However, this contribution vanishes \cite{Buchbinder:2015qsa} so it is omitted here.}
\begin{equation}
	\langle L^{\abar}_{\a}(z_{1}) L^{\bbar}_{\b}(z_{2}) L^{\cbar}_{\g}(z_{3}) \rangle = f^{\abar \bbar \cbar} \, \frac{\boldsymbol{x}_{13 \a}{}^{\a'} \boldsymbol{x}_{23 \b}{}^{\b'}}{(\boldsymbol{x}_{13}^{2})^{2} (\boldsymbol{x}_{23}^{2})^{2}} \; \cH_{\a' \b' \g}(\boldsymbol{X}_{3},\Q_{3}) \, ,
	\label{100}
\end{equation}
The correlation function is required to satisfy the following properties:
\begin{enumerate}
	\item[\textbf{(i)}] \textbf{Scaling constraint:}
	
	Under scale transformations the correlation function must transform as
	\begin{equation}
	\langle L^{\abar}_{\a}(z'_{1}) L^{\bbar}_{\b}(z'_{2}) L^{\cbar}_{\g}(z'_{3}) \rangle = (\l^{2})^{9/2} \langle L^{\abar}_{\a}(z_{1}) L^{\bbar}_{\b}(z_{2}) L^{\cbar}_{\g}(z_{3}) \rangle \, , 
	\end{equation}
	which gives rise to the homogeneity constraint on $\cH$:
	\begin{equation}
	\cH_{\a \b \g}(\l^{2} \boldsymbol{X}, \l \Q) = (\l^{2})^{-3/2} \cH_{\a \b \g}(\boldsymbol{X}, \Q) \, . \label{N=1 LLL - scaling constraint}
	\end{equation}
	\item[\textbf{(ii)}] \textbf{Differential constraints:}
	
	The conservation equation for the flavour current results in
	\begin{equation}
	D^{\a}_{(1)} \langle L^{\abar}_{\a}(z_{1}) L^{\bbar}_{\b}(z_{2}) L^{\cbar}_{\g}(z_{3}) \rangle = 0 \, .
	\end{equation}
	Using identities \eqref{Three-point building blocks 1c - differential identities 3}, \eqref{Three-point building blocks 1c - differential identities 4}, we obtain a differential constraint on $\cH$:
	\begin{equation}
		\cD^{\a} \cH_{\a \b \g}(\boldsymbol{X}, \Q) = 0 \, . \label{N=1 LLL - differential constraint 1}
	\end{equation}
	We need not consider the conservation law at $z_{2}$ as we can use an algebraic constraint instead.

	\item[\textbf{(iii)}] \textbf{Point permutation symmetry:}
	
	The symmetry under permutation of points ($z_{1}$ and $z_{2}$) results in the following constraint on the correlation function:
	\begin{equation}
		\langle L^{\abar}_{\a}(z_{1}) L^{\bbar}_{\b}(z_{2}) L^{\cbar}_{\g}(z_{3}) \rangle = - \langle L^{\bbar}_{\b}(z_{2}) L^{\abar}_{\a}(z_{1}) L^{\cbar}_{\g}(z_{3}) \rangle \, ,
	\end{equation}
	which constrains the tensor $\cH$ so that
	\begin{equation}
		\cH_{\a \b \g}(\boldsymbol{X},\Q) = \cH_{\b \a \g}(-\boldsymbol{X}^{\text{T}}, - \Q) \, . \label{N=1 LLL - point switch identity 1}
	\end{equation}
	On the other hand the symmetry under permutation of points $z_{1}$ and $z_{3}$ results in
	\begin{equation}
	\langle L^{\abar}_{\a}(z_{1}) L^{\bbar}_{\b}(z_{2}) L^{\cbar}_{\g}(z_{3}) \rangle = - \langle L^{\cbar}_{\g}(z_{3}) L^{\bbar}_{\b}(z_{2}) L^{\abar}_{\a}(z_{1}) \rangle \, ,
	\end{equation}
	which gives rise to the point-switch identity
	\begin{align}
	\cH_{\a \b \g}(\boldsymbol{X}_{3} , \Q_{3}) = \frac{ \boldsymbol{x}_{13 \g}^{\g'} (\boldsymbol{x}_{13}^{-1})_{\a}{}^{\a'} \boldsymbol{x}_{13}^{ \b' \s} \boldsymbol{X}_{3 \s \b}}{\boldsymbol{X}_{3}^{4} \boldsymbol{x}_{13}^{4}} \, \cH_{\g' \b' \a'}(- \boldsymbol{X}_{1}^{\text{T}}, - \Q_{1}) \, . \label{N=1 LLL - point switch identity 2}
	\end{align}
\end{enumerate}

To solve this problem systematically let's decompose the tensor $\cH$ into irreducible components:
\begin{equation}
	\cH_{\a \b \g }(\boldsymbol{X},\Q) = \sum_{i} c_{i} \, \cH_{i \; \a \b \g}(\boldsymbol{X},\Q) \, .
\end{equation}
It is also more convenient to work with $X_{m}$ instead of $\boldsymbol{X}_{\a \b}$. We have
\begin{subequations}
	\begin{align}
		\cH_{1 \; \a \b \g} &= \ve_{\a \b} \Q_{\g} A(X) \, , \\
		\cH_{2 \; \a \b \g} &= \ve_{\a \b} (\g^{a})_{\g}{}^{\d} \Q_{\d} B_{a}(X) \, , \\
		\cH_{3 \; \a \b \g} &= (\g^{a})_{\a \b} \Q_{\g} C_{a}(X) \, , \\
		\cH_{4 \; \a \b \g} &= (\g^{a})_{\a \b} (\g^{b})_{\g}{}^{\d} \Q_{\d} D_{a b}(X) \, .
	\end{align}
\end{subequations}
Here we have used the fact that every matrix anti-symmetric in $\a,\b$ is proportional to $\ve_{\a \b}$, every matrix symmetric in $\a,\b$ is proportional to a gamma-matrix, and that since $\cH$ is Grassmann odd it follows that $\cH$ is linear in $\Q$ due to $\Q_{\a} \Q_{\b} \Q_{\g} = 0$. 
Due to the scaling property~\eqref{N=1 LLL - scaling constraint} it follows that the functions $A, B, C, D$ have dimension $-2$. 
From eq.~\eqref{N=1 LLL - point switch identity 1} it also follows that 
\begin{subequations}
\begin{align}
	& \hspace{6mm} A (X) = A(-X) \, , \hspace{10mm} B_{a} (X) = B_{a} (-X) \, , \\
	&C_{a} (X)= - C_{a} (-X) \, , \hspace{10mm} D_{ab} (X)= - D_{ab} (-X) \, .  \label{new20}
\end{align}
\end{subequations}
It is easy to see that the conservation equation \eqref{N=1 LLL - differential constraint 1} splits into the two independent equations
\begin{subequations}
	\begin{align}
		&\partial^{\a} \cH_{\a \b \g} = 0 \, , \label{N=1 LLL - differential constraint 1a} \\[2mm]
		&(\g^{t})^{\a \t} \Q_{\t} \partial_{t} \cH_{\a \b \g} = 0 \, . \label{N=1 LLL - differential constraint 1b}
	\end{align}
\end{subequations}
Imposing~\eqref{N=1 LLL - differential constraint 1a} results in the algebraic equations
\begin{equation}
	A(X) = - D^{a}{}_{a}(X) \, , \hspace{10mm} C_{a}(X) = B_{a}(X) + \e_{a}{}^{mn} D_{mn}(X) \, .
\label{new111}
\end{equation}
While on the other hand from~\eqref{N=1 LLL - differential constraint 1b} we obtain 
\begin{subequations}
	\begin{align}
		&\partial^{a} \big\{ B_{a}(X) + C_{a}(X) - \e_{a}{}^{mn} D_{mn}(X) \big\} = 0 \, , \label{new111a}\\[2mm]
		&\partial_{t} A(X) + \e_{t}{}^{ma} \partial_{m} B_{a}(X) - \e_{t}{}^{ma} \partial_{m} C_{a}(X) \label{new111b} \\
		& \hspace{16mm} - \partial^{m} D_{mt}(X) + \partial_{t} D^{a}{}_{a}(X) - \partial^{m} D_{tm}(X) = 0 \nonumber  \, .
	\end{align}
\end{subequations}
Using eqs.~\eqref{new111},~\eqref{new111a},~\eqref{new111b} we obtain that $B_a$ and $D_{ab}$ satisfy 
\be 
\pa^a B_{a}(X)=0 \,, \hspace{10mm} \pa^a D_{ab}(X)=0\,. 
\label{new111c}
\ee
Therefore we see that this problem is reduced to finding transverse tensors $B_a$ and $D_{ab}$ of dimension $-2$ satisfying~\eqref{new20}. The tensors
$A$ and $C$ are then found using eq.~\eqref{new111}. It is not difficult to show that the solution to this problem is given by 
\begin{subequations}
	\begin{align}
		A(X) &= 0 \, , \hspace{20mm} B_{a}(X) = 0 \, , \\[1mm]
		C_{a}(X) &= \frac{X_{a}}{X^{3}} \, , \hspace{15mm} D_{ab}(X) = \e_{a b c} \frac{X^{c}}{X^{3}} \, ,
	\end{align}
\end{subequations}
with $c_{3} = - 2 c_{4}$. Hence this correlation function is fixed up to a single real coefficient which we denote $d_{\cN=1}$.
Converting back to spinor notation we find\footnote{Note that since $\Q_{\a} \Q_{\b} \Q_{\g} = 0$ we can replace $\boldsymbol{X}$ 
with $X$ in~\eqref{new10}.}
\begin{equation}
	\cH_{\a \b \g}(\boldsymbol{X},\Q) = \frac{\text{i} d_{\cN=1}}{\boldsymbol{X}^{3}} \Big\{  \boldsymbol{X}_{\a \b} \Q_{\g} - \ve_{\a \g} \boldsymbol{X}_{\b}{}^{\d} \Q_{\d} - \ve_{\b \g} \boldsymbol{X}_{\a}{}^{\d} \Q_{\d}  \Big\} \, .
\label{new10}	
\end{equation}
One may also check that this solution satisfies the point-switch identity~\eqref{N=1 LLL - point switch identity 2}. This agrees with the result in \cite{Buchbinder:2015qsa}, which 
was computed in a different way. Our method has the advantage that it systematically takes care of all possible irreducible 
components of $\cH$, hence it is more useful when $\cH$ is a tensor of high rank.


\section{Mixed correlators in \texorpdfstring{$\cN=1$}{N=1} superconformal field theory}\label{section4}


\subsection{The correlation function \texorpdfstring{$\langle L J L \rangle$}{< L J L >}}


Let us first consider the correlation function $\langle L^{\bar{a}}_{\a}(z_{1}) J_{\g_{1} \g_{2} \g_{3}}(z_{2}) L^{\bar{b}}_{\b}(z_{3}) \rangle$. 
Using the general expression~\eqref{Three-point function - general ansatz}, 
it has the form
%
\begin{equation}
	\langle L^{\abar}_{\a}(z_{1}) J_{\g_{1} \g_{2} \g_{3}}(z_{2}) L^{\bbar}_{\b}(z_{3}) \rangle = \frac{ \d^{\abar \bbar} \, \hat{\boldsymbol{x}}_{13 \a}{}^{\a'} \hat{\boldsymbol{x}}_{23 (\g_{1}}{}^{\g_{1}'} \hat{\boldsymbol{x}}_{23 \g_{2}}{}^{\g_{2}'} \hat{\boldsymbol{x}}_{23 \g_{3})}{}^{\g_{3}'} }{(\boldsymbol{x}_{13}^{2})^{3/2} (\boldsymbol{x}_{23}^{2})^{5/2}} \, \cH_{\a' \b , \g_{1}' \g_{2}' \g_{3}'}(\boldsymbol{X}_{3}, \Q_{3}) \, , 
\end{equation}
where $\cH$ is totally symmetric in three of its indices, $\cH_{\a \b , \g_{1} \g_{2} \g_{3}} = \cH_{\a \b , (\g_{1} \g_{2} \g_{3})}$. The correlation function is also required to satisfy:
\begin{enumerate}
	\item[\textbf{(i)}] \textbf{Scaling constraint:}
	
	Under scale transformations the correlation function transforms as
	\begin{equation}
		\langle L^{\bar{a}}_{\a}(z'_{1}) J_{\g_{1} \g_{2} \g_{3}}(z'_{2}) L^{\bar{b}}_{\b}(z'_{3}) \rangle = (\l^{2})^{11/2} \langle L^{\bar{a}}_{\a}(z_{1}) J_{\g_{1} \g_{2} \g_{3}}(z_{2}) L^{\bar{b}}_{\b}(z_{3}) \rangle \, , 
	\end{equation}
	which implies that we have the following homogeneity constraint on $\cH$:
	\begin{equation}
		\cH_{\a \b , \g_{1} \g_{2} \g_{3}}(\l^{2} \boldsymbol{X}, \l \Q) = (\l^{2})^{-5/2} \cH_{\a \b , \g_{1} \g_{2} \g_{3}}(\boldsymbol{X}, \Q) \, . \label{N=1 LJL - scaling constraint}
	\end{equation}
	\item[\textbf{(ii)}] \textbf{Differential constraints:}
	
	The differential constraints on the flavour current and supercurrent result in the following constraints on the correlation function:
	\begin{subequations}
			\begin{align}
			D^{\a}_{(1)} \langle L^{\bar{a}}_{\a}(z_{1}) J_{\g_{1} \g_{2} \g_{3}}(z_{2}) L^{\bar{b}}_{\b}(z_{3}) \rangle = 0 \, , \\[2mm]  
			D^{\g_{1}}_{(2)} \langle L^{\bar{a}}_{\a}(z_{1}) J_{\g_{1} \g_{2} \g_{3}}(z_{2}) L^{\bar{b}}_{\b}(z_{3}) \rangle = 0 \, .
		\end{align}
	\end{subequations}
Using identities \eqref{Three-point building blocks 1c - differential identities 3} and \eqref{Three-point building blocks 1c - differential identities 4}, these result in the following differential constraints on $\cH$:
\begin{subequations}
	\begin{align}
		\cD^{\a} \cH_{\a \b , \g_{1} \g_{2} \g_{3}}(\boldsymbol{X}, \Q) &= 0 \, , \label{N=1 LJL - differential constraint 1}\\[2mm]
		\cQ^{\g_{1}} \cH_{\a \b , \g_{1} \g_{2} \g_{3}}(\boldsymbol{X}, \Q)  &= 0 \, . \label{N=1 LJL - differential constraint 2}
	\end{align}
\end{subequations}

\item[\textbf{(iii)}] \textbf{Point permutation symmetry:}
	
	The symmetry under permutation of points ($z_{1}$ and $z_{3}$) results in the following constraint on the correlation function:
	\begin{equation}
		\langle L^{\bar{a}}_{\a}(z_{1}) J_{\g_{1} \g_{2} \g_{3}}(z_{2}) L^{\bar{b}}_{\b}(z_{3}) \rangle = - \langle L^{\bar{b}}_{\b}(z_{3}) J_{\g_{1} \g_{2} \g_{3}}(z_{2}) L^{\bar{a}}_{\a}(z_{1}) \rangle \, ,
	\end{equation}
	which results in the point-switch identity
	\begin{align}
	\begin{split}
		\cH_{\a \b , \g_{1} \g_{2} \g_{3}}(\boldsymbol{X}_{3}, \Q_{3}) &= - \frac{ \boldsymbol{x}_{13}{}^{\b'}{}_{\b}  (\boldsymbol{x}_{13}^{-1})_{\a}{}^{\a'} \boldsymbol{x}_{13}^{\g_{1}' \d_{1}} \boldsymbol{X}_{3 \d_{1} \g_{1}} \boldsymbol{x}_{13}^{\g_{2}' \d_{2}} \boldsymbol{X}_{3 \d_{2} \g_{2}} \boldsymbol{x}_{13}^{\g_{3}' \d_{3}} \boldsymbol{X}_{3 \d_{3} \g_{3}}}{\boldsymbol{X}_{3}^{8} \boldsymbol{x}_{13}^{8}} \\[2mm]
		& \hspace{8mm} \times \cH_{\b' \a' , \g_{1}' \g_{2}' \g_{3}'}(- \boldsymbol{X}_{1}^{\text{T}}, - \Q_{1}) \, . \label{N=1 LJL - point switch identity}
	\end{split}
	\end{align}
\end{enumerate}
Thus we need to solve for the tensor $\cH$ subject to the constraints \eqref{N=1 LJL - scaling constraint}, 
\eqref{N=1 LJL - differential constraint 1}, \eqref{N=1 LJL - differential constraint 2} and \eqref{N=1 LJL - point switch identity}. We begin by combining two of the three $\g$ indices into a vector index, while simultaneously imposing a $\g$-trace constraint to remove the component 
anti-symmetric in $\g_{1} , \g_{2}$,
\begin{equation}
	\cH_{\a \b , \g_{1} \g_{2} \g_{3}} = (\g^{m})_{\g_{2} \g_{3}} \cH_{\a \b , \g_{1} m} \, , \hspace{5mm} (\g^{m})^{\t \g} \cH_{\a \b , \g m} = 0 \, . \label{N=1 LJL - gamma trace constraint}
\end{equation}
Since our correlator is Grassmann odd the function $\cH_{\a \b, \g m}$ must be linear in $\Q$. 
Just like the flavour current three-point function, linearity in $\Q$ implies that the differential constraints \eqref{N=1 LJL - differential constraint 1} 
and \eqref{N=1 LJL - differential constraint 2} are respectively equivalent to
\begin{subequations}
	\begin{align}
		\partial^{\a} \cH_{\a \b , \g m} &= 0 \, ,  \hspace{8mm} (\g^{t})^{\a \t} \Q_{\t} \partial_{t} \cH_{\a \b , \g m} = 0 \, , \label{N=1 LJL - differential constraint 1a}\\[1mm]
		\partial^{\g} \cH_{\a \b , \g m} &= 0 \, ,  \hspace{8mm} (\g^{t})^{\g \t} \Q_{\t} \partial_{t} \cH_{\a \b , \g m} = 0 \, . \label{N=1 LJL - differential constraint 2a}
	\end{align}
\end{subequations}
Now let us decompose $\cH$ into irreducible components
\begin{equation}
	\cH_{\a \b , \g m} = \sum_{i} c_{i} \, \cH_{i \; \a \b , \g m} \, ,
\end{equation}
where
\begin{subequations}
	\begin{align}
		\cH_{1 \; \a \b , \g m} &= \ve_{\a \b} \Q_{\g} A_{m}(X) \, , \\
		\cH_{2 \; \a \b , \g m} &= \ve_{\a \b} (\g^{a})_{\g}{}^{\d} \Q_{\d} B_{ma}(X) \, , \\
		\cH_{3 \; \a \b , \g m} &= (\g^{a})_{\a \b} \Q_{\g} C_{ma}(X) \, , \\
		\cH_{4 \; \a \b , \g m} &= (\g^{a})_{\a \b} (\g^{b})_{\g}{}^{\d} \Q_{\d} D_{mab}(X) \, .
	\end{align}
\end{subequations}
It follows from eq.~\eqref{N=1 LJL - scaling constraint} that the dimension of $A, B, C, D$ is $-3$. 
We now impose the differential constraints \eqref{N=1 LJL - differential constraint 1a} and \eqref{N=1 LJL - differential constraint 2a}, along with the gamma-trace constraint \eqref{N=1 LJL - gamma trace constraint}. After imposing \eqref{N=1 LJL - differential constraint 1a}, \eqref{N=1 LJL - differential constraint 2a} the terms $O(\Q^{0})$ imply
\begin{subequations}
	\begin{align}
	& \hspace{15mm} A_{m}(X) = 0 \, , \hspace{10mm} C_{mn}(X) = 0 \, , \\[2mm]
	&B_{ma}(X) = - \e_{nra} D_{mnr}(X) \, , \hspace{8mm} \eta^{na} D_{mna}(X) = 0 \, ,
	\end{align}
\end{subequations}
while the terms $O(\Q^{2})$ give the differential constraints
\begin{subequations}
	\begin{align}
		&\pa^{t} B_{mt}(X) = 0 \, , \\[1mm]
		&\pa^{t} D_{mnt}(X) = 0 \, , \\[1mm]
		&\pa^{t} \big\{ B_{mt}(X) + \e_{t}{}^{an} D_{mna}(X) \big\} = 0 \, , \\[1mm]
		&\pa^{t} \big\{ D_{mnt}(X) + D_{mtn}(X) - \eta_{tn} D_{m}{}^{a}{}_{a}(X) + \e_{nt}{}^{a} B_{ma}(X) \big\} = 0 \, .
	\end{align}
\end{subequations}
Imposing the gamma-trace condition \eqref{N=1 LJL - gamma trace constraint} results in
\begin{subequations}
	\begin{align}
	\eta^{ma} B_{ma}(X) &= 0 \, , \hspace{8mm} \e^{qma} B_{ma}(X) = 0 \, , \\[1mm]
	\eta^{ma} D_{mna}(X) &=0 \, , \hspace{8mm} \e^{qma} D_{mna}(X) = 0 \, .
	\end{align}
\end{subequations}
One may show that the differential and algebraic constraints above are mutually consistent and reduce to:
\begin{subequations}
	\begin{align}
		\pa^{t} B_{mt}(X) &= 0 \, , \hspace{8mm} \pa^{t} D_{mnt}(X) = 0 \, , \\[1mm]
		\eta^{na} D_{mna}(X) &= 0 \, ,  \hspace{8mm} \eta^{ma} D_{mna}(X) =0 \, , \\[1mm]
		& \hspace{-10mm} B_{ma}(X) = - \e_{nra} D_{mnr}(X) \, ,
	\end{align}
\end{subequations}
where $B_{ma}$ is symmetric and traceless, $D_{mna}$ is symmetric in the first and last index. After some calculation one can show that general solutions consistent with the scaling property \eqref{N=1 LJL - scaling constraint} and the above constraints is
\begin{align}
	B_{ma}(X) &= \frac{\eta_{ma}}{X^{3}} - \frac{3 X_{m} X_{a}}{X^{5}} \, , \label{new21} \\[1mm]
	D_{mna}(X) &= \e_{ndm} \frac{X^{d} X_{a}}{X^{5}} + \e_{nda} \frac{X^{d} X_{m}}{X^{5}} \label{new22} \, ,
\end{align}
with $c_{2} = c_{4}$. Hence, the three-point correlation function is determined up to a single free parameter which we denote $c_{\cN=1}$. 
Our solution is then
\begin{equation}
\langle L^{\abar}_{\a}(z_{1}) J_{\g_{1} \g_{2} \g_{3}}(z_{2}) L^{\bbar}_{\b}(z_{3}) \rangle = \frac{ \d^{\abar \bbar} \, \boldsymbol{x}_{13 \a}{}^{\a'} \boldsymbol{x}_{23 (\g_{1}}{}^{\g_{1}'} \boldsymbol{x}_{23 \g_{2}}{}^{\g_{2}'} \boldsymbol{x}_{23 \g_{3})}{}^{\g_{3}'} }{(\boldsymbol{x}_{13}^{2})^{2} (\boldsymbol{x}_{23}^{2})^{4}} \, \cH_{\a' \b , \g_{1}' \g_{2}' \g_{3}'}(\boldsymbol{X}_{3}, \Q_{3}) \, ,
\label{newnew1} 
\end{equation}
where
\begin{equation}
	\cH_{\a \b , \g_{1} \g_{2} \g_{3}}(\boldsymbol{X},\Q) = (\g^{m})_{\g_{2} \g_{3}} \cH_{\a \b , \g_{1} m}(\boldsymbol{X},\Q) \, ,
	\label{newnew2}
\end{equation}
\begin{align}
	\cH_{\a \b , \g m}(\boldsymbol{X},\Q) = \text{i} c_{\cN=1} (\g^{a})_{\g}{}^{\d} \Q_{\d} \, \Big\{  \ve_{\a \b} \, B_{ma}(X) + (\g^{n})_{\a \b} \, D_{mna}(X) \Big\} \, ,
	\label{newnew3}
\end{align}
with $B$ and $D$ given in eqs.~\eqref{new21}, \eqref{new22}. In spinor notation, this is equivalent to
\begin{align}
	\cH_{\a \b , \g_{1} \g_{2} \g_{3}}(\boldsymbol{X},\Q) &= \text{i} c_{\cN=1} \, \bigg\{ \frac{\ve_{\a \b}}{\boldsymbol{X}^{3}} \big( \ve_{\g_{1} \g_{2}} \Q_{\g_{3}} + \ve_{\g_{1} \g_{3}} \Q_{\g_{2}} \big) + \frac{1}{\boldsymbol{X}^{5}} \big( \ve_{ \g_{2} \a} \boldsymbol{X}_{\b \g_{3}} \boldsymbol{X}_{\g_{1}}{}^{\d} \Q_{\d} 
	\label{N=1 mixed correlator solution}\\[2mm]
	& \hspace{5mm} + \ve_{ \g_{2} \b} \boldsymbol{X}_{\a \g_{3}} \boldsymbol{X}_{\g_{1}}{}^{\d} \Q_{\d} + \ve_{ \g_{1} \a} \boldsymbol{X}_{\g_{2} \g_{3}} \boldsymbol{X}_{\b}{}^{\d} \Q_{\d} + \ve_{ \g_{1} \b} \boldsymbol{X}_{\g_{2} \g_{3}} \boldsymbol{X}_{\a}{}^{\d} \Q_{\d} \nonumber \\[1mm] 
	& \hspace{7mm} - \ve_{ \g_{2} \g_{3}} \boldsymbol{X}_{\a \b} \boldsymbol{X}_{\g_{1}}{}^{\d} \Q_{\d} - \boldsymbol{X}_{\g_{2} \g_{3}} \boldsymbol{X}_{\a \b} \Q_{\g_{1}} - 3 \ve_{\a \b} \boldsymbol{X}_{\g_{2} \g_{3}} \boldsymbol{X}_{\g_{1}}{}^{\d} \Q_{\d} \big) \bigg\} \, . \nonumber
\end{align}
Finally, one must check that this solution also satisfies the point-switch identity. With the aid of identities \eqref{Three-point building blocks 1a - properties 1}, \eqref{Three-point building blocks 1a - properties 2}, it is a relatively straightforward exercise to show that the point-switch 
identity~\eqref{N=1 LJL - point switch identity} is indeed satisfied. 


\subsection{The correlation function \texorpdfstring{$\langle J J L \rangle$}{< J J L >}}\label{subsection 4.2}


Let us now discuss the remaining mixed correlation function  
\begin{equation}
	\langle J_{\b_{1} \b_{2} \b_{3}}(z_{1}) J_{\g_{1} \g_{2} \g_{3}}(z_{2}) L_{\a}(z_{3}) \rangle \, .
	\label{new26}
\end{equation} 
Here the correlator can exist only if the flavour group contains $U(1)$-factors, so will assume that the flavour group is just $U(1)$.
At the component level this correlation function contains $\langle T_{ab} (x_{1}) T_{mn}(x_{2}) L_{c}(x_{3}) \rangle$, which was shown to vanish 
in any conformal field theory after imposing all differential constraints and symmetries \cite{Giombi:2011rz}. 
As we will show, the same occurs in the supersymmetric theory. However, we will see that~\eqref{new26} vanishes 
without needing to impose the conservation equation for $L_{\a}(z_{3})$.
The general expression for this correlation function is
\begin{align}
	\langle J_{\b_{1} \b_{2} \b_{3}}(z_{1}) J_{\g_{1} \g_{2} \g_{3}}(z_{2}) L_{\a}(z_{3}) \rangle &= \frac{ \hat{\boldsymbol{x}}_{13 (\b_{1}}{}^{\b_{1}'} \hat{\boldsymbol{x}}_{13 \b_{2}}{}^{\b_{2}'} \hat{\boldsymbol{x}}_{13 \b_{3})}{}^{\b_{3}'} \hat{\boldsymbol{x}}_{23 (\g_{1}}{}^{\g_{1}'} \hat{\boldsymbol{x}}_{23 \g_{2}}{}^{\g_{2}'} \hat{\boldsymbol{x}}_{23 \g_{3})}{}^{\g_{3}'} }{(\boldsymbol{x}_{13}^{2})^{5/2} (\boldsymbol{x}_{23}^{2})^{5/2}} \\
	& \hspace{5mm} \times \cH_{ \b_{1}' \b_{2}' \b_{3}' \g_{1}' \g_{2}' \g_{3}' \a}(\boldsymbol{X}_{3}, \Q_{3}) \, , \nonumber
\end{align}
where $\cH$ has the symmetry property $\cH_{\b_{1} \b_{2} \b_{3} \g_{1} \g_{2} \g_{3} \a} = \cH_{(\b_{1} \b_{2} \b_{3}) (\g_{1} \g_{2} \g_{3}) \a}$.  
The correlation function is required to satisfy:
\begin{enumerate}
	\item[\textbf{(i)}] \textbf{Scaling constraint:}
	
	Under scale transformations it transforms as
	\begin{equation}
		\langle J_{\b_{1} \b_{2} \b_{3}}(z_{1}') J_{\g_{1} \g_{2} \g_{3}}(z_{2}') L_{\a}(z_{3}') \rangle = (\l^{2})^{13/2} \langle J_{\b_{1} \b_{2} \b_{3}}(z_{1}) J_{\g_{1} \g_{2} \g_{3}}(z_{2}) L_{\a}(z_{3}) \rangle \, , 
	\end{equation}
	which results in the constraint
	\begin{equation}
		\cH_{\b_{1} \b_{2} \b_{3} \g_{1} \g_{2} \g_{3} \a}( \l^{2} \boldsymbol{X} , \l \Q) = (\l^{2})^{-7/2} \cH_{\b_{1} \b_{2} \b_{3} \g_{1} \g_{2} \g_{3} \a}( \boldsymbol{X} , \Q) \, . \label{N=1 JJL - scaling constraint}
	\end{equation}
	\item[\textbf{(ii)}] \textbf{Differential constraint:}
	
	The conservation law on the supercurrent implies
	\begin{equation}
		D^{\b_{1}}_{(1)} \langle J_{\b_{1} \b_{2} \b_{3}}(z_{1}) J_{\g_{1} \g_{2} \g_{3}}(z_{2}) L_{\a}(z_{3}) \rangle = 0 \, ,
	\end{equation}
	which results in a differential constraint on $\cH$:
	\begin{equation}
		\cD^{\b_{1}} \cH_{\b_{1} \b_{2} \b_{3} \g_{1} \g_{2} \g_{3} \a}( \boldsymbol{X} , \Q) = 0 \, . \label{N=1 JJL - differential constraint 1}
	\end{equation}

	\item[\textbf{(iii)}] \textbf{Point permutation symmetry:}
	
	The symmetry under permutation of points $z_{1}$ and $z_{2}$ implies the following constraint on the correlation function:
	\begin{equation}
		\langle J_{\b_{1} \b_{2} \b_{3}}(z_{1}) J_{\g_{1} \g_{2} \g_{3}}(z_{2}) L_{\a}(z_{3}) \rangle = - \langle J_{\g_{1} \g_{2} \g_{3}}(z_{2}) J_{\b_{1} \b_{2} \b_{3}}(z_{1}) L_{\a}(z_{3}) \rangle \, ,
	\end{equation}
	which results in the identity
	\begin{equation}
		\cH_{\b_{1} \b_{2} \b_{3} \g_{1} \g_{2} \g_{3} \a}( \boldsymbol{X} , \Q) = - \cH_{\g_{1} \g_{2} \g_{3} \b_{1} \b_{2} \b_{3} \a}( - \boldsymbol{X}^{\text{T}} , - \Q) \, . \label{N=1 JJL - point switch identity}
	\end{equation}
\end{enumerate}
Thus, we need to solve for the tensor $\cH$ subject to the constraints \eqref{N=1 JJL - scaling constraint}, \eqref{N=1 JJL - differential constraint 1} and \eqref{N=1 JJL - point switch identity}. Note that we also must impose one more differential constraint 
\begin{equation}
D^{\a}_{(3)} \langle J_{\b_{1} \b_{2} \b_{3}}(z_{1}) J_{\g_{1} \g_{2} \g_{3}}(z_{2}) L_{\a}(z_{3}) \rangle = 0 \, ,
\label{new29}
\end{equation}
which is quite non-trivial in this formalism. Fortunately, constraints~\eqref{N=1 JJL - scaling constraint}, \eqref{N=1 JJL - differential constraint 1} and \eqref{N=1 JJL - point switch identity} are sufficient to show that correlator~\eqref{new26} vanishes, hence we will not need to consider~\eqref{new29}.

To start, we combine two of the three $\b$, $\g$ indices into a vector index, and impose $\g$-trace constraints to remove antisymmetric components
\begin{equation}
	\cH_{\b_{1} \b_{2} \b_{3} \g_{1} \g_{2} \g_{3} \a}( \boldsymbol{X} , \Q) = (\g^{a})_{\b_{2} \b_{3}} (\g^{b})_{\g_{2} \g_{3}} \cH_{\b_{1} a , \g_{1} b , \a}( \boldsymbol{X} , \Q) \, ,
\end{equation}
\begin{equation}
	 (\g^{a})^{\t \b} \cH_{\b a , \g b , \a}( \boldsymbol{X} , \Q) = 0 \, , \hspace{8mm}  (\g^{b})^{\t \g} \cH_{\b a , \g b , \a}( \boldsymbol{X} , \Q) = 0 \, . 
\label{N=1 JJL - gamma trace constraint}
\end{equation}
Now let us split $\cH$ into symmetric and antisymmetric parts in the first and second pair of indices
\begin{equation}
	\cH_{\b a , \g b , \a} = \cH_{(\b a , \g b) , \a} + \cH_{[\b a , \g b] , \a} \,.
\end{equation}
Due to the symmetry properties,~\eqref{N=1 JJL - point switch identity} implies that $\cH_{(\s a , \g b) , \a}$
is an even function of $X$,~\footnote{As in the previous case, our correlator is Grassmann odd which means we can replace $\boldsymbol{X}$ with $X$.} 
while $\cH_{[\b a , \g b] , \a}$ is odd. Therefore they do not mix in the conservation law \eqref{N=1 JJL - differential constraint 1} 
and may be considered independently. In irreducible components, $\cH_{(\b a , \g b) , \a}$ has the decomposition
\begin{equation}
	\cH_{(\b a , \g b) , \a} = \sum_{i} \, \cH_{i \; (\b a , \g b) , \a} \, ,
\end{equation}
where
\begin{subequations}
	\begin{align}
		\cH_{1 \; (\b a , \g b) , \a} &= \ve_{\b \g} \Q_{\a} A_{[ab]}(X) \, , \\
		\cH_{2 \; (\b a , \g b) , \a} &= \ve_{\b \g} (\g^{m})_{\a}{}^{\d} \Q_{\d} B_{m[ab]}(X) \, , \\
		\cH_{3 \; (\b a , \g b) , \a} &= (\g^{m})_{\b \g} \Q_{\a} C_{m(ab)}(X) \, , \\
		\cH_{4 \; (\b a , \g b) , \a} &= (\g^{m})_{\b \g} (\g^{n})_{\a}{}^{\d} \Q_{\d} D_{mn(ab)}(X) \, .
	\end{align}
\end{subequations}
Here we have made explicit the algebraic symmetry properties of $A,B,C$ and $D$, which by virtue of \eqref{N=1 JJL - point switch identity} are all even functions of $X$. Now due to linearity in $\Q$, the differential constraint \eqref{N=1 JJL - differential constraint 1} is equivalent to the pair of equations
\begin{equation}
	\pa^{\b} \cH_{\b a , \g b , \a} = 0 \, ,  \hspace{8mm} (\g^{t})^{\b \t} \Q_{\t} \pa_{t} \cH_{\b a , \g b , \a} = 0 \, . \label{N=1 JJL - differential constraint 1a}
\end{equation}
After imposing \eqref{N=1 JJL - differential constraint 1a}, the terms $O(\Q^{0})$ imply
\begin{subequations}
	\begin{align}
		& \hspace{6mm} A_{m[ab]}(X) = 0 \, , \hspace{8mm} B_{m[ab]}(X) = 0 \, , \\[2mm]
		& \hspace{8mm} C_{m(ab)}(X) + \e_{m}{}^{rs} D_{rs(ab)}(X) = 0 \, , \label{N=1 JJL - constraints 1} \\[2mm]
		& \eta^{mn} D_{mn(ab)}(X) = 0 \, , \hspace{5mm} \eta^{ma} D_{mn(ab)}(X) = 0 \, , \label{N=1 JJL - constraints 2}
	\end{align}
\end{subequations}
so $\cH_{1 \; (\b a , \g b) , \a} = \cH_{2 \; (\b a , \g b) , \a} = 0$. The terms $O(\Q^{2})$ then result in the differential constraints
\begin{subequations}
	\begin{align}
		& \hspace{12mm} \pa^{m} \big\{ -C_{m(ab)}(X) + \e_{m}{}^{rs} D_{rs(ab)}(X) \big\} = 0 \, , \label{N=1 JJL - constraints 3} \\[1mm]
		& \e_{c}{}^{tm} \pa_{t} C_{m(ab)}(X) - \pa^{m} D_{mc(ab)}(X) - \pa^{m} D_{cm(ab)}(X) = 0 \, . \label{N=1 JJL - constraints 4}
	\end{align}
\end{subequations}
Imposing the gamma-trace condition \eqref{N=1 JJL - gamma trace constraint} results in
\begin{subequations}
	\begin{align}
		\eta^{ma} C_{m(ab)}(X) &= 0 \, , \hspace{8mm} \e_{c}{}^{ma} C_{m(ab)}(X) = 0 \, , \label{N=1 JJL - constraints 5} \\[1mm]
		\eta^{ma} D_{mn(ab)}(X) &=0 \, , \hspace{8mm} \e_{c}{}^{ma} D_{mn(ab)}(X) = 0 \, . \label{N=1 JJL - constraints 6}
	\end{align}
\end{subequations}
Altogether \eqref{N=1 JJL - constraints 1}, \eqref{N=1 JJL - constraints 3} and \eqref{N=1 JJL - constraints 5} imply that $C$ is a totally symmetric, 
traceless, transverse and even function of $X$. Let's try to construct such a tensor by analysing its irreducible components. 
To determine which irreducible components are permitted, let us trade each vector index for a pair of spinor indices. 
Since $C$ is completely symmetric and traceless, it is equivalent to $C_{(\a_{1} \, .... \, \a_{6})}$. In addition since $C$ is even in $X_{\a \b}$ only 
irreducible structures (that is, totally symmetric tensors) of rank $4$ and $0$ in $X_{\a \b}$ can contribute to the solution. 
Going back to vector indices, let us denote these components of $C$ as $C_{1 \, (mn)}(X)$ and $C_{2}(X)$. 

Since it is not possible to construct a rank three tensor $C_{(mnk)}$ out of $C_{1 \, (mn)}(X)$ and $C_{2}(X)$,
the tensor $C_{mnk}$ vanishes. Hence, $\cH_{3 \; (\b a , \g b) , \a} = 0$.

Given this information, the remaining set of equations imply that $D$ is now a totally symmetric, traceless and transverse tensor that is even in $X$. 
Following a similar argument, the symmetries imply that it has irreducible components $D_{1 \, (mnab)}(X)$, $D_{2 \, (mn)}(X)$ and $D_{3}(X)$. 
We are now equipped with enough information to construct an explicit solution for $D$. Using the symmetries and the scaling property \eqref{N=1 JJL - scaling constraint} we have the most general ansatz
\begin{align}
	\begin{split}
		D_{(mnab)}(X) &= \frac{d_{1}}{X^{4}} \big[ \eta_{ma} \eta_{nb} + \eta_{mb} \eta_{na} + \eta_{mn} \eta_{ab} \big] \\
		& \hspace{5mm} + \frac{d_{2}}{X^{6}} \big[ \eta_{mn} X_{a} X_{b} + \eta_{ma} X_{n} X_{b} + \eta_{mb} X_{a} X_{n} \\[-2mm]
		& \hspace{18mm} + \eta_{na} X_{m} X_{b} + \eta_{nb} X_{m} X_{a} + \eta_{ab} X_{m} X_{n} \big] + \frac{d_{3}}{X^{8}} X_{m} X_{n} X_{a} X_{b} \, .
	\end{split}
\end{align} 
Requiring that $D$ be traceless and transverse fixes all the $d_{i}$ to $0$. Hence, $D=0$, and $\cH_{(\b a , \g b) , \a}$ vanishes. 

In a similar way we consider $\cH_{[\b a , \g b] , \a}$ for which we have the following decomposition 
\begin{equation}
	\cH_{[\b a , \g b] , \a} = \sum_{i} \, \cH_{i \; [\b a , \g b] , \a} \, ,
	\label{new100}
\end{equation}
where
\begin{subequations}
	\begin{align}
		\cH_{1 \; [\b a , \g b] , \a} &= \ve_{\b \g} \Q_{\a} A_{(ab)}(X) \, , \label{new101}\\
		\cH_{2 \; [\b a , \g b] , \a} &= \ve_{\b \g} (\g^{m})_{\a}{}^{\d} \Q_{\d} B_{m(ab)}(X) \, , \label{new102} \\
		\cH_{3 \; [\b a , \g b] , \a} &= (\g^{m})_{\b \g} \Q_{\a} C_{m[ab]}(X) \, ,  \label{new103} \\
		\cH_{4 \; [\b a , \g b] , \a} &= (\g^{m})_{\b \g} (\g^{n})_{\a}{}^{\d} \Q_{\d} D_{mn[ab]}(X) \label{new104} \, .
	\end{align}
\end{subequations}
In this case, $A, B, C, D$ are now odd functions in $X$. Imposing the conservation equations and vanishing of the $\g$-trace we obtain
the following set of constraints:
\begin{subequations}
\begin{align}
& A_{(ab)}(X) =0\,, \qquad B_{m(ab)}(X)=0 \, \label{new105} \\
& D^m_{ \ m [ab]}(X) =0 \,, \qquad C_{m [ab] } (X)+ \e_{m}{}^{rs} D_{rs [a b]} (X)=0\,, \label{new106} \\
& C^m_{ \ [m b]}(X) =0\,, \qquad D^m_{\ n [m b]}(X) =0 \,, \label{new107} \\
& \e^{c ma} C_{m [ab]}(X)=0\,, \label{new107.1} \\
& \e^{c ma} D_{mn [ab]}(X)=0\,. \label{new107.2}
\end{align}
\end{subequations}
We see that the functions $A$ and $B$ vanish. To show that $C_{m [ab]}$ vanishes we consider eq.~\eqref{new107.1}
and use the fact that in three dimensions an antisymmetric tensor is equivalent to a vector
\be 
C_{m [ab]}(X)= \e_{ab}{}^{q} \tilde{C}_{mq}(X) \,. 
\label{new108}
\ee
Hence from~\eqref{new107.1} it follows that 
\be
\tilde{C}_{ab}(X) - \eta_{ab} \tilde{C}^d_{\ d}(X) =0\,. 
\label{new109}
\ee
Contracting with $\eta^{ab}$ we find that $\tilde{C}^d_{\ d}=0$, hence $\tilde{C}_{ab}=0$. It also implies that $C_{m [ab]}$=0. 
In a similar way using eq.~\eqref{new107.2} one can show that $D_{mn [ab]}=0$. This means that $\cH_{[\b a , \g b] , \a} = 0$.
Hence the three-point function of two supercurrents and one flavour 
current~\eqref{new26} vanishes.


\section{Comments on the absence of parity violating structures}\label{section5}


In \cite{Giombi:2011rz} it was shown that correlation functions of conserved current in three-dimensional conformal field theories can have parity violating structures. Specifically, it was defined as follows. Given a conserved current 
\be
J_{\a_{1} \a_{2} \dots \a_{2s-1} \a_{2s}} (x) = (\g^{m_{1}})_{\a_{1} \a_{2}} \dots (\g^{m_{s}})_{\a_{2s-1} \a_{2s}} 
J_{m_{1} \dots m_{s}}(x) \, ,
\label{5.1}
\ee
we can construct 
\be 
J_s (x, \l)= J_{\a_1 \a_2 \dots \a_{2s-1} \a_{2s}} (x) \, \l^{\a_1} \dots  \l^{\a_{2s}} \,, 
\label{5.2}
\ee
where $\l^{\a}$ are auxiliary commuting spinors. The action of parity is then $x \to -x$, $\l \to {\rm i} \l$.
In theories with a parity symmetry, $J_{\mu_1 \dots \mu_s}(x) $ acquires a sign $(-1)^s$ under parity
and $J_s (x, \l)$ is invariant. However, as was shown in \cite{Giombi:2011rz} correlation functions admit contributions 
which are odd under parity. 
In particular, it was shown that a parity odd contribution to the mixed correlator 
of the energy-momentum tensor $T_{mn}$ and two flavour currents $L_k^{\bar a}$ can arise. 
Translating their result into our notation it can be written as follows
\begin{equation}
	\langle T_{m n} (x_{1}) L^{\abar}_{k}(x_{2}) L^{\bbar}_{p}(x_{3}) \rangle_{odd} = \frac{\d^{\abar \bbar}}{x_{13}^{3} x_{12}^{3} x_{23}} \, 
	\cI_{mn, m'n'}(x_{13}) \, I_{kk'}(x_{23}) \, t_{m' n' k' p}(X_{3}) \, ,
	\label{5.3}
\end{equation}
where 
\begin{equation}
	t_{m n k p}(X) = \e_{npq} \frac{X^{q} X_{m} X_{k}}{X^{3}} + \e_{nkq} \frac{X^{q} X_{m} X_{p}}{X^{3}} \, .
	\label{5.4}
\end{equation}
Here the $X_{i}$ are the three-point building blocks introduced by Osborn and Petkou in \cite{Osborn:1993cr}, while the objects $I_{mn}(x)$ and $\cI_{mn, m'n'}(x)$ are the inversion tensors acting on vectors and rank-2 tensors respectively. They are defined as follows
\begin{equation}
	I_{mn}(x) = \eta_{m n} - 2 \frac{x_{m} x_{n}}{x^{2}} \, ,
	\label{5.5}
\end{equation}
\begin{equation}
	\cI_{mn, m'n'}(x) = \frac{1}{2} \big\{ I_{m m'}(x) I_{n n'}(x) + I_{m n'}(x) I_{n m'}(x) \big\} - \frac{1}{3} \eta_{m m'} \eta_{n n'} \, .
	\label{5.6}
\end{equation}
An important and specific feature of all parity violating terms is appearance of the $\e$-tensor. 

In $\cN=1$ supersymmetric theories the supercurrent $J_{\a \b\g}$ and the flavour current multiplet $L_{\a}^{\bar a}$ contain the following 
conserved currents
\begin{subequations}
\begin{align}
& T_{\a \b \g \d}  = D_{(\d} J_{\a \b\g)}|\,, \quad T_{\a \b \g \d}=(\g^{m})_{(\a \b}  (\g^{n})_{\g \d)} T_{mn}\,, \quad \pa^m T_{mn}=0\,, \quad 
\eta^{mn}T_{mn}=0\,,
 \label{5.7} \\
& Q_{\a \b \g} = J_{\a \b\g}|\,, \qquad Q_{\a \b \g} = (\g^{m})_{\a \b} Q_{m \g}\,, \quad \pa^m Q_{m \a}=0\,, \quad 
(\g^{m})^{\a \b} Q_{m \a}=0\,,
\label{5.8} \\
& V_{\a \b}^{\bar a} = D_{(\a} L_{\b)}^{\bar a}|\,, \qquad V^{\abar}_{\a \b}=(\g^{m})_{\a \b}  V_{m}^{\bar a}\,, \quad \pa^m V_{m}^{\bar a}=0\,, 
 \label{5.9}
\end{align}
\end{subequations}
where $T_{mn}$ is the energy-momentum tensor, $Q_{m \g}$ is the supersymmetry current and $V_{m}^{\bar a}$ is a vector current. 
Hence, the mixed correlators studied in the previous section give rise to the following correlators in terms of components
\be
\langle T_{mn} (x_1) T_{pq} (x_2) V_k (x_3) \rangle \,, \quad 
\langle Q_{m \a} (x_1) Q_{n \b} (x_2) V_k (x_3) \rangle \,, \quad
\langle V_k^{\bar a} (x_1) T_{mn} (x_2) V_p^{\bar b} (x_3)\rangle\,. 
\label{5.10}
\ee
The first two correlators vanish because the entire superspace correlator~\eqref{new26} vanishes. The last one is, in general, non-zero 
and fixed up to one overall coefficient. It can be computed using eqs.~\eqref{newnew1}, \eqref{N=1 mixed correlator solution} using the superspace reduction procedure
\bea
&&
\langle V_k^{\bar a} (x_1) T_{mn} (x_2) V_p^{\bar b} (x_3)\rangle= 
\label{5.11} \\
&&
\frac{1}{16}( \g_k)^{\a_1 \a_2} ( \g_m)^{(\b_1 \b_2} ( \g_n)^{\b_3 \b_4)} ( \g_p)^{\g_1 \g_2} 
D_{(1)\a_1} D_{(2)\b_1} D_{(3)\g_1} 
\langle L_{\a_2}^{\bar a} (z_1) J_{\b_2 \b_3 \b_4} (z_2) L_{\g_2}^{\bar b} (z_3)\rangle \big| \,.
\nonumber 
\eea
Here the bar-projection denotes setting the fermionic coordinates $\q_{\a}$ to zero. We will not perform the reduction explicitly, instead we will indirectly determine whether~\eqref{5.11} is even or odd under parity. For this it is sufficient to study whether or not the $\e$-tensor appears upon reduction. Since 
\be 
\e_{mnp} =\frac{1}{2} {\rm tr} (\g_m \g_n \g_p) \, ,
\label{5.12}
\ee
it is enough to count the number of gamma-matrices: If the number of $\g$-matrices appearing in the superspace reduction is 
even the $\e$-tensor cannot arise and the contribution is parity even, if the number of $\g$-matrices is odd
the contribution is parity odd. Let us perform the counting. Since in~\eqref{5.11} we act with just three covariant derivatives 
before setting all $\theta_{i}=0$ (where $i=1, 2, 3$ is the index labelling the three point) only term linear and cubic 
in $\theta_{i}$ will contribute. Let us concentrate on the terms linear on $\theta_i$. Since the function $\cH$ 
in~\eqref{N=1 mixed correlator solution}  is already linear in $\theta_i$ we can set $\theta_i=0$ in $\boldsymbol{x}_{ij}$ and $\boldsymbol{X}$.
This makes $\boldsymbol{x}_{ij}$ and $\boldsymbol{X}$ symmetric and proportional to a gamma-matrix. 
Now we have four gamma-matrices in~\eqref{5.11}, four gamma-matrices coming from $\boldsymbol{x}_{ij}$  in eq.~\eqref{newnew1}, 
zero or two gamma-matrices coming from $\cH$ in~\eqref{N=1 mixed correlator solution} and also one more gamma-matrix 
contained in $\Theta_3$, see eq.~\eqref{Three-point building blocks 1c}. Overall we have odd number of gamma-matrices at this point. 
However, superspace covariant derivatives also contain gamma-matrices, see eq.~\eqref{Covariant spinor derivatives}. 
Since we are considering terms linear in $\theta_i$ and setting $\theta_i=0$ upon differentiating it is easy to realise that 
in the three derivatives $D_{(1)\a_1} D_{(2)\b_1} D_{(3)\g_1} $ we must take one derivative with respect to $x_i$ 
and two derivatives with respect to $\theta_i$. This gives one more gamma-matrix making the total number even. 
Terms cubic in $\theta$ can be considered in a similar way. They also yield an even number of gamma-matrices.
Hence, the entire contribution~\eqref{5.11} is parity even.

In a similar way we can count the number of gamma-matrices in the superspace reduction of~\eqref{100}, \eqref{new10}:
\bea
&&
\langle V_m^{\bar a} (x_1) V_n^{\bar b} (x_2) V_k^{\bar c} (x_3)\rangle= 
\label{5.13} \\
&&
-\frac{1}{8}( \g_m)^{\a_1 \a_2} ( \g_n)^{\b_1 \b_2}  ( \g_k)^{\g_1 \g_2} 
D_{(1)\a_1} D_{(2)\b_1} D_{(3)\g_1} 
\langle L_{\a_2}^{\bar a} (z_1) L_{\b_2}^{\bar a} (z_2)  L_{\g_2}^{\bar b} (z_3)\rangle \big| \,.
\nonumber 
\eea
An analysis similar to the above shows that this contribution is also parity even. Finally, one can also consider the superspace 
reduction of the three-point function of the supercurrent
\bea
&&
\langle T_{mn} (x_1) T_{k \ell} (x_2)  T_{pq} (x_3) \rangle = 
\frac{1}{64}( \g_m)^{(\a_1 \a_2}( \g_n)^{\a_3 \a_4)}  ( \g_k)^{(\b_1 \b_2} ( \g_{\ell})^{\b_3 \b_4)} ( \g_p)^{(\g_1 \g_2} ( \g_q)^{\g_3 \g_4)} 
\nonumber  \\
&&
D_{(1)\a_1} D_{(2)\b_1} D_{(3)\g_1} 
\langle J_{\a_2 \a_3 \a_4} (z_1) J_{\b_2 \b_3 \b_4} (z_2)J_{\g_2 \g_3 \g_4} (z_3)\rangle \big| \, ,
\label{5.14} 
\eea
and 
\bea
&&
\langle T_{mn} (x_1) Q_{k \b} (x_2)  Q_{p \g} (x_3) \rangle = 
\nonumber  \\
&&
 \frac{1}{16}( \g_m)^{(\a_1 \a_2}( \g_n)^{\a_3 \a_4)}  ( \g_k)^{\b_1 \b_2}( \g_p)^{\g_1 \g_2}
D_{(1)\a_1} 
\langle J_{\a_2 \a_3 \a_4} (z_1) J_{\b \b_1 \b_2} (z_2)J_{\g \g_1 \g_2} (z_3)\rangle \big| \,.
\label{5.15} 
\eea
The three-point function of the supercurrent was found in~\cite{Buchbinder:2015qsa}. We will not repeat it here since the 
expression for it is quite long. However, a similar analysis shows that the contributions~\eqref{5.14} and~\eqref{5.15}
are parity even.\footnote{In general, if a superspace three-point function is fixed up to an overall coefficient 
it is expected to be parity even because this contribution is expected to exist in a free theory of a real scalar superfield.}

This means that no parity violating structures can arise in three-point functions of $T_{mn}, Q_{m \a}$ and $V_m^{\bar a}$ 
in superconformal field theories. Maldacena and Zhiboedov proved in~\cite{Maldacena:2011jn} that if a three-dimensional conformal 
field theory possesses a higher spin conserved current then it is essentially a free theory. 
Since a free theory has only parity even contributions to the three-point functions of conserved currents, the correlators involving one or more higher spin conserved currents admit 
only parity even structures. 
This leads us to conclude that $\cN=1$ supersymmetry forbids parity violating structures in all three-point functions of conserved currents
unless the assumptions of the Maldacena--Zhiboedov theorem are violated. 
The strongest assumption of the theorem is that the theory under consideration contains unique conserved current of spin two which is 
the energy-momentum tensor. Some properties of theories possessing more than one conserved current with spin two were discussed in \cite{Maldacena:2011jn}. In supersymmetric theory the energy-momentum tensor is a component of the supercurrent. 
One can also consider a different supermultiplet containing a conserved spin two current,
namely 
\be 
J_{(\a_1 \a_2 \a_3 \a_4)}\,, \qquad D^{\a_1} J_{(\a_1 \a_2 \a_3 \a_4)}=0\,. 
\label{5.16}
\ee
The lowest component of $J_{(\a_1 \a_2 \a_3 \a_4)}$ is a conserved spin two current which is not the energy-momentum tensor. 
Note that $J_{(\a_1 \a_2 \a_3 \a_4)}$ also contains a conserved higher-spin current. It will be interesting 
to perform a systematic study of three-point functions of $J_{(\a_1 \a_2 \a_3 \a_4)}$ to see if they allow any parity violating structures.


\section{Mixed correlators in \texorpdfstring{$\cN=2$}{N=2} superconformal field theory}\label{section6}


Now we will generalise our method to mixed three-point functions in superconformal field theory with 
$\cN=2$ supersymmetry. A specific feature of three-dimensional $\cN=2$ superconformal field theories is contact terms in correlation functions 
of the conserved currents~\cite{Closset:2012vg, Closset:2012vp}. In this paper, 
we study correlation functions at non-coincident points where the contact terms do not contribute.


\subsection{Supercurrent and flavour current multiplets}

The 3D, $\cN=2$ supercurrent was studied in~\cite{Zupnik:1988en, Dumitrescu:2011iu, Kuzenko:2011rd, Kuzenko:2012qg}.
It is a primary, dimension $2$ symmetric spin-tensor $J_{\a \b}$, which obeys the conservation equation
\begin{equation}
	D^{I\a} J_{\a \b} = 0 \, , 
\end{equation}
and has the following superconformal transformation law:
\begin{equation}
	\d J_{\a \b} = - \x J_{\a \b} - 2 \s(z) J_{\a \b} + 2 \l(z)^{\g}_{(\a} J_{\b ) \g} \, .
\end{equation}
The general formalism in section 2 allows the two-point function to be determined up to a single real coefficient
\begin{equation}
	\langle J_{\a \b}(z_{1}) J^{\a' \b'}(z_{2}) \rangle = b_{\cN=2} \frac{\boldsymbol{x}_{12 (\a}{}^{\a'} \boldsymbol{x}_{12 \b)}{}^{\b'} }{(\boldsymbol{x}_{12}^{2})^{3}} \,.
\end{equation}
It's then a simple exercise to show that the two-point function has the right symmetry properties under permutation of superspace points
\begin{equation}
	\langle J_{\a \b}(z_{1}) J_{\a' \b'}(z_{2}) \rangle = \langle J_{\a' \b'}(z_{2}) J_{\a \b}(z_{1}) \rangle \, ,
\end{equation}
and also satisfies the conservation equation
\begin{equation}
	D_{(1)}^{I \a} \langle J_{\a \b}(z_{1}) J_{\a' \b'}(z_{2}) \rangle = 0 \, , \hspace{10mm} z_{1} \neq z_{2} \, .
\end{equation}

Similarly, the 3D $\cN=2$ flavour current is a primary, dimension $1$ scalar superfield $L$, which obeys the conservation equation
\begin{equation}
	\big( D^{\a (I} D^{J)}_{\a} - \frac{1}{2} \d^{IJ} D^{\a K} D^{K}_{\a} \big) L = 0 \, , 
\end{equation}
and transforms under the superconformal group as
\begin{equation}
	\d L = -\x L - \s(z) L \, .
\end{equation}
As in the $\cN=1$ case, we assume the $\cN=2$ superconformal field theory in question has a set of flavour currents $L^{\abar}$ associated with a simple flavour group. Due to the absence of spinor or $R$-symmetry indices, the $\cN=2$ flavour current two-point function is fixed up to a single real coefficient $a_{\cN=2}$ as follows
\begin{equation}
	\langle L^{\abar}(z_{1}) L^{\bbar}(z_{2}) \rangle = a_{\cN=2} \frac{\d^{\abar \bbar}  }{\boldsymbol{x}_{12}^{2}} \, .
\end{equation}
The two-point function obeys the correct symmetry properties under permutation of superspace points, $\langle L^{\abar}(z_{1}) L^{\bbar}(z_{2}) \rangle = \langle L^{\bbar}(z_{2}) L^{\abar}(z_{1}) \rangle$, and also satisfies the conservation equation 
\begin{equation}
	\big( D^{\a (I}_{(1)} D^{J)}_{(1)\a} - \frac{1}{2} \d^{IJ} D^{\a K}_{(1)} D^{K}_{(1)\a} \big) \langle L^{\abar}(z_{1}) L^{\bbar}(z_{2}) \rangle = 0 \, , \hspace{5mm} z_{1} \neq z_{2} \, .
\end{equation}
In the next section we will compute the mixed correlation functions associated with the $\cN=2$ supercurrent and flavour current multiplets. There are two possibilities to consider, they are 
\begin{equation}
	\langle L^{\abar}(z_{1}) J_{\a \b}(z_{2}) L^{\bbar}(z_{3}) \rangle   \, , \hspace{10mm}   \langle J_{\a \b}(z_{1}) J_{\g \d}(z_{2}) L(z_{3}) \rangle \, .
\end{equation}
Note that in second case we are considering a $U(1)$ flavour current.


\subsection{The correlation function \texorpdfstring{$\langle L J L \rangle$}{< L J L >}}\label{subsection 6.2}

First let us consider the $\langle L J L \rangle$ case first. Using the general ansatz, we have
\begin{equation}
	\langle L^{\abar}(z_{1}) J_{\a \b}(z_{2}) L^{\bbar}(z_{3}) \rangle = \frac{\d^{\abar \bbar} \boldsymbol{x}_{23 \a}{}^{\a'} \boldsymbol{x}_{23 \b}{}^{\b'}}{ \boldsymbol{x}_{13}^{2} (\boldsymbol{x}_{23}^{2})^{3}} \; \cH_{\a' \b'}(\boldsymbol{X}_{3}, \Q_{3}) \, ,
\end{equation}
where $\cH_{\a \b} = \cH_{(\a \b)}$. The correlation function is also required to satisfy the following:

\begin{enumerate}
	\item[\textbf{(i)}] \textbf{Scaling constraint:}
	
	Under scale transformations the correlation function transforms as
	\begin{equation}
		\langle L^{\abar}(z_{1}') J_{\a \b}(z_{2}') L^{\bbar}(z_{3}') \rangle = (\l^{2})^{4} \langle L^{\abar}(z_{1}) J_{\a \b}(z_{2}) L^{\bbar}(z_{3}) \rangle \, ,
	\end{equation}
	from which we find the homogeneity constraint
	\begin{equation}
		\cH_{\a \b}(\l^{2} \boldsymbol{X}, \l \Q) = (\l^{2})^{-2} \cH_{\a \b }(\boldsymbol{X}, \Q) \, . \label{N=2 LJL - scaling constraint}
	\end{equation}
	
	\item[\textbf{(ii)}] \textbf{Differential constraints:}
	
	The differential constraints on the flavour current and supercurrent result in the following constraints on the correlation function:
	\begin{subequations}
		\begin{align}
			\big(D^{\s (I}_{(1)} D^{J)}_{(1)\s} - \frac{1}{2} \d^{IJ} D^{\s K}_{(1)} D^{K}_{(1)\s} \big) \langle L^{\abar}(z_{1}) J_{\a \b}(z_{2}) L^{\bbar}(z_{3}) \rangle &= 0 \, ,  \\[2mm]
			D^{I\a}_{(2)} \langle L^{\abar}(z_{1}) J_{\a \b}(z_{2}) L^{\bbar}(z_{3}) \rangle = 0 \, . \hspace{10mm} &
		\end{align}
	\end{subequations}
	These result in the following differential constraints on $\cH$:
	\begin{subequations}
		\begin{align}
			(\cD^{\s (I} \cD^{J)}_{\s} - \frac{1}{2} \d^{IJ} \cD^{\s K} \cD^{K}_{\s}) \cH_{\a \b}(\boldsymbol{X}, \Q) &= 0 \, ,  \label{N=2 LJL - differential constraint 1} \\[2mm]
			\cQ^{I \a} \cH_{\a \b}(\boldsymbol{X}, \Q) = 0 \, . \hspace{10mm} \label{N=2 LJL - differential constraint 2} &
		\end{align}
	\end{subequations}

	\item[\textbf{(iii)}] \textbf{Point permutation symmetry:}
	
	The symmetry under permutation of points ($z_{1}$ and $z_{3}$) results in the following constraint on the correlation function:
	\begin{equation}
		\langle L^{\abar}(z_{1}) J_{\a \b}(z_{2}) L^{\bbar}(z_{3}) \rangle = \langle L^{\bbar}(z_{3}) J_{\a \b}(z_{2}) L^{\abar}(z_{1}) \rangle \, ,
	\end{equation}
	which results in the point-switch identity
	\begin{align}
		\cH_{\a \b}(\boldsymbol{X}_{3}, \Q_{3}) = \frac{\boldsymbol{x}_{13}^{\s \s'} \boldsymbol{X}_{3 \s' \a} \boldsymbol{x}_{13}^{\r \r'} \boldsymbol{X}_{3 \r' \b} }{\boldsymbol{X}_{3}^{6} \, \boldsymbol{x}_{13}^{6}} \; \cH_{\s \r}(-\boldsymbol{X}_{1}^{\text{T}}, -\Q_{1}) \, . \label{N=2 LJL - point switch identity}
	\end{align}	
\end{enumerate}
The symmetry properties of $\cH$ allow us to trade the spinor indices for a vector index
\begin{equation}
	\cH_{\a \b}(\boldsymbol{X}, \Q) = (\g^{m})_{\a \b} \, \cH_{m}(\boldsymbol{X},\Q) \, .
\end{equation}
The most general expansion for $\cH_{m}(\boldsymbol{X},\Q)$ is then
\begin{equation}
	\cH_{m}(\boldsymbol{X},\Q) = A_{m}(X) - \frac{\text{i}}{2} \Q^{2} B_{m}(X) + (\Q \Q )^{n} C_{mn}(X) + \frac{1}{8} \Q^{4} D_{m}(X) \, ,
\end{equation}
where we have defined
\begin{equation}
	(\Q \Q)_{m} = - \frac{1}{2} (\g_{m})^{\a \b} (\Q \Q)_{\a \b} \, , \hspace{5mm} (\Q \Q)_{\a \b} = \Q_{\a}^{I} \Q_{\b}^{J} \ve_{IJ} \, ,
\end{equation}
and accounted for the $\cN=2$ identity
\begin{equation}
	\Q^{2} \Q_{\a}^{I} \Q_{\b}^{J} \ve_{IJ} = 0 \, . \label{N=2 identity}
\end{equation}
The prefactors in front of $B$ and $D$ have been chosen for convenience, and as in the $\cN=1$ case it is more convenient to work with $X^{m}$ instead of $\boldsymbol{X}^{\a \b}$. Imposing \eqref{N=2 LJL - differential constraint 2} results in the differential constraints
\begin{subequations} \label{N=2 LJL - differential constraints 2a}
	\begin{align}
		& \pa^{m} A_{m}(X) = 0 \, , \\
		& \pa^{m} B_{m}(X) = 0 \, , \\
		& \e^{mnt} \pa_{n} C_{mt}(X) = 0 \, , \\
		& B_{q}(X) + \e_{qmn} \pa^{n} A^{m}(X) = 0 \, , \\
		& D_{q}(X) - \e_{q m n} \pa^{n} B^{m}(X) = 0 \, , \\
		& \pa^{m} \big\{ C_{mt}(X) + C_{tm}(X) - \eta_{mt} C^{a}{}_{a}(X) \big\} = 0 \, ,
	\end{align}
\end{subequations}
and the algebraic constraints
\begin{subequations}
	\begin{align}
		& C^{a}{}_{a}(X) = 0 \, , \\
		& \e^{qmt} C_{mt}(X) = 0 \, ,
	\end{align}
\end{subequations}
which imply that $C$ is symmetric and traceless. Furthermore the scaling condition \eqref{N=2 LJL - scaling constraint} allows us to construct the solutions
\begin{subequations}
	\begin{align}
		& A_{m}(X) = a \, \frac{X_{m}}{X^{3}} \, , \\[1mm]
		& B_{m}(X) = b \, \frac{X_{m}}{X^{4}} \, , \\[1mm]
		& C_{mn}(X) = c \, \bigg( \frac{\eta_{mn}}{X^{3}} - \frac{3 X_{m} X_{n} }{X^{5}} \bigg) \, , \\[1mm]
		& D_{m}(X) = d \, \frac{X_{m}}{X^{5}} \, .
	\end{align}
\end{subequations}
Together \eqref{N=2 LJL - differential constraints 2a} imply $B_{m}(X) = D_{m}(X) = 0$, while $a$ and $c$ remain as two free parameters. Hence the solution for $\cH$ becomes
\begin{equation}
	\cH_{\a \b}(\boldsymbol{X},\Q) = \tilde{c}_{\cN=2} \frac{X_{ \a \b}}{X^{3}} + {\rm i} c_{\cN=2} \, \bigg\{ \frac{\Q^{I}_{\a} \Q^{J}_{\b} \ve_{IJ} }{X^{3}} + \frac{3}{2} \frac{X_{\a \b} X^{\g \d} \Q^{I}_{\g} \Q^{J}_{\d} \ve_{IJ} }{X^{5}}\bigg\} \, .
	\label{1000}
\end{equation}
After some lengthy calculation it turns out that only the second structure satisfies the conservation 
equation \eqref{N=2 LJL - differential constraint 1}. Hence there is only one linearly independent structure 
in the correlation function that is compatible with the differential constraints. Therefore we find that the final solution is
\begin{equation}
	\langle L^{\abar}(z_{1}) J_{\a \b}(z_{2}) L^{\bbar}(z_{3}) \rangle = \frac{\d^{\abar \bbar} \boldsymbol{x}_{23 \a}{}^{\a'} \boldsymbol{x}_{23 \b}{}^{\b'}}{ \boldsymbol{x}_{13}^{2} (\boldsymbol{x}_{23}^{2})^{3}} \; \cH_{\a' \b'}(\boldsymbol{X}_{3}, \Q_{3}) \, ,
\end{equation}
with
%
\begin{align}
	\cH_{\a \b}(\boldsymbol{X},\Q) = \text{i} c_{\cN=2} \, \bigg\{ \frac{\Q^{I}_{\a} \Q^{J}_{\b} \ve_{IJ} }{\boldsymbol{X}^{3}} + \frac{3}{2} \frac{\boldsymbol{X}_{\a \b} \boldsymbol{X}^{\g \d} \Q^{I}_{\g} \Q^{J}_{\d} \ve_{IJ}}{\boldsymbol{X}^{5}} \bigg\} \, .
\end{align}
In deriving this result, we Taylor expanded the denominator in~\eqref{1000}
using $X^{2} = \boldsymbol{X}^{2} - \frac{1}{4} \Q^{4}$, which follows 
from~\eqref{Three-point building blocks 2}, \eqref{Three-point building blocks 1a - properties 3}, 
and then used the $\cN=2$ identity \eqref{N=2 identity}. 
It may also be shown that this structure satisfies the point-switch identity \eqref{N=2 LJL - point switch identity}.

The supercurrent $J_{\a \b}$ leads to the following $\cN=1$ supermultiplets (here the bar-projections denotes setting 
$\theta^{I=\mathbf{2}}$ to zero and $D^{\a}= D^{\a, I= \mathbf{1}}$):\footnote{From here we will use bold $R$-symmetry indices to distinguish them from other types of indices.}
\begin{subequations}
	\begin{align}
		& S_{\a \b}= J_{\a \b}|\,,\qquad  \qquad \ D^{\a} S_{\a \b}=0\,, \label{6.1} \\
		& J_{\a \b \g}= {\rm i} D_{(\a}^{\mathbf{2}} J_{\b \g)}\,, \qquad D^{\a} J_{\a \b \g}=0\,. \label{6.2}
	\end{align}
\end{subequations}
In these equations $J_{\a \b \g}$ is the $\cN=1$ supercurrent and $S_{\a \b}$ is the additional $\cN=1$ supermultiplet 
containing the second supersymmetry current and the $R$-symmetry current. 
Similarly, the $\cN=2$ flavour current leads to 
\begin{subequations}
	\begin{align}
		& S= L^{\bar a}|\,,  \label{6.3} \\
		& L_{\a }^{\bar a}= {\rm i} D_{\a}^{\mathbf{2}} L^{\bar a}\,, \qquad D^{\a} L_{\a }^{\bar a}=0\,, \label{6.4}
	\end{align}
\end{subequations}
where $L_{\a}^{\bar a}$ is the $\cN=1$ flavour current and $S$ is unconstrained. Hence, the 
$\cN=2$ three-point function $\langle L^{\abar}(z_{1}) J_{\a \b}(z_{2}) L^{\bbar}(z_{3}) \rangle $ 
contains three-point functions of the following conserved component currents: the energy-momentum tensor, conserved vector currents, the supersymmetry currents and the $R$-symmetry current. 
All these three-point functions can be found by superspace reduction and are fixed by the $\cN=2$ superconformal symmetry 
up to one overall coefficient (or vanish). A simple gamma-matrix counting procedure similar to the one discussed in the previous section shows that all these correlators are parity even. 


\subsection{The correlation function \texorpdfstring{$\langle J J L \rangle$}{< J J L >}}

For this example, the general ansatz gives
\begin{equation}
	\langle J_{\a \b}(z_{1}) J_{\g \d}(z_{2}) L(z_{3}) \rangle = \frac{\boldsymbol{x}_{13 \a}{}^{\a'} \boldsymbol{x}_{13 \b}{}^{\b'} \boldsymbol{x}_{23 \g}{}^{\g'} \boldsymbol{x}_{23 \d}{}^{\d'} }{ (\boldsymbol{x}_{13}^{2})^{3} (\boldsymbol{x}_{23}^{2})^{3}} \; \cH_{\a' \b' \g' \d'}(\boldsymbol{X}_{3}, \Q_{3}) \, ,
\end{equation}
where $\cH_{\a \b \g \d} = \cH_{(\a \b) (\g \d)}$. The correlation function is required to satisfy the following:

\begin{enumerate}
	\item[\textbf{(i)}] \textbf{Scaling constraint:}
	
	Under scale transformations the correlation function transforms as
	\begin{equation}
		\langle J_{\a \b}(z_{1}') J_{\g \d}(z_{2}') L(z_{3}') \rangle = (\l^{2})^{5} \langle J_{\a \b}(z_{1}) J_{\g \d}(z_{2}) L(z_{3}) \rangle \, ,
	\end{equation}
	from which we find the homogeneity constraint
	\begin{equation}
		\cH_{\a \b \g \d}(\l^{2} \boldsymbol{X}, \l \Q) = (\l^{2})^{-3} \cH_{\a \b \g \d }(\boldsymbol{X}, \Q) \, . \label{N=2 JJL - scaling constraint}
	\end{equation}
	
	\item[\textbf{(ii)}] \textbf{Differential constraints:}
	
	The differential constraints on the flavour current and supercurrent result in the following constraints on the correlation function:
	\begin{subequations}
		\begin{align}
			D^{I\a}_{(1)} \langle J_{\a \b}(z_{1}) J_{\g \d}(z_{2}) L(z_{3}) \rangle = 0 \, . \hspace{12mm} & \\[2mm]
			\big(D^{\s (I}_{(3)} D^{J)}_{(3)\s} - \frac{1}{2} \d^{IJ} D^{\s K}_{(3)} D^{K}_{(3)\s} \big) \langle J_{\a \b}(z_{1}) J_{\g \d}(z_{2}) L(z_{3}) \rangle &= 0 \, . \label{N=2 - JJL - differential constraint 2}
		\end{align}
	\end{subequations}
	The first equation results in the following differential constraints on $\cH$:
	\begin{equation}
		\cD^{I \a} \cH_{\a \b \g \d}(\boldsymbol{X}, \Q) = 0 \, . \hspace{10mm} \label{N=2 JJL - differential constraint 1} 
	\end{equation}
	The second constraint \eqref{N=2 - JJL - differential constraint 2} is more difficult to handle in this formalism, however we will demonstrate how to deal with it later.

	\item[\textbf{(iii)}] \textbf{Point permutation symmetry:}
	
	The symmetry under permutation of points $z_{1}$ and $z_{2}$ results in the following constraint on the correlation function:
	\begin{equation}
		\langle J_{\a \b}(z_{1}) J_{\g \d}(z_{2}) L(z_{3}) \rangle = \langle J_{\g \d}(z_{2}) J_{\a \b}(z_{1}) L(z_{3}) \rangle \, ,
	\end{equation}
	which results in the point-switch identity
	\begin{align}
		\cH_{\a \b \g \d}(\boldsymbol{X}, \Q) = \cH_{\g \d \a \b }(-\boldsymbol{X}^{\text{T}}, -\Q) \, . \label{N=2 JJL - point switch identity}
	\end{align}
	
\end{enumerate}
Now due to the symmetry properties of $\cH$, we may trade pairs of symmetric spinor indices for vector indices
\begin{equation}
	\cH_{(\a \b) (\g \d)}(\boldsymbol{X}, \Q) = (\g^{m})_{\a \b} (\g^{n})_{\g \d}\, \cH_{mn}(\boldsymbol{X},\Q) \, .
\end{equation}
Now if we split $\cH_{mn}$ into symmetric and anti-symmetric parts
\begin{align}
	\cH_{mn}(\boldsymbol{X},\Q) &= \cH_{(mn)}(\boldsymbol{X},\Q) + \cH_{[mn]}(\boldsymbol{X},\Q) \nonumber \\[2mm]
	&= \cH_{(mn)}(\boldsymbol{X},\Q) + \e_{mnt} \cH^{t}(\boldsymbol{X},\Q) \, ,
\end{align}
then the point-switch identity implies
\begin{equation}
	\cH_{(mn)}(\boldsymbol{X},\Q) = \cH_{(mn)}(-\boldsymbol{X}^{\text{T}},-\Q) \, , \hspace{8mm} \cH_{t}(\boldsymbol{X},\Q) = -\cH_{t}(-\boldsymbol{X}^{\text{T}},-\Q) \, .
\end{equation}
General expansions consistent with the index structure and symmetries are
\begin{subequations}
	\begin{align}
		\cH_{(mn)}(\boldsymbol{X},\Q) &= A_{(mn)}(X) + \Q^{2} B_{(mn)}(X) + (\Q \Q )^{s} C_{(mn)s}(X) + \Q^{4} D_{(mn)}(X) \, , \\[2mm]
		& \hspace{-4mm} \cH_{t}(\boldsymbol{X},\Q) = A_{t}(X) + \Q^{2} B_{t}(X) + (\Q \Q )^{s} C_{ts}(X) + \Q^{4} D_{t}(X) \, .
	\end{align}
\end{subequations}
All the tensors comprising $\cH_{(mn)}$ are even functions of $X$, while those in the expansion for $\cH_{t}$ are odd functions of $X$. Furthermore, due to symmetry arguments the tensors $\cH_{(mn)}$ and $\cH_{t}$ do not mix in the conservation law \eqref{N=2 JJL - differential constraint 1}, hence they may be considered independently. First let us analyse $\cH_{(mn)}$; imposing \eqref{N=2 LJL - differential constraint 1} results in the differential constraints
\begin{subequations} \label{N=2 JJL - differential constraints 1a}
	\begin{align}
		& \pa^{m} A_{(mn)}(X) = 0 \, , \label{N=2 JJL - differential constraints 1a - 1} \\
		& \pa^{m} B_{(mn)}(X) = 0 \, , \label{N=2 JJL - differential constraints 1a - 2} \\
		& \e^{mrs} \pa_{r} C_{(mn)s}(X) = 0 \, , \\
		& 2 B_{(qn)}(X) + \text{i} \e_{q}{}^{mt} \pa_{t} A_{(mn)}(X) = 0 \, , \label{N=2 JJL - differential constraints 1a - 3} \\
		& 4 D_{(qn)}(X) + \text{i} \e_{q}{}^{mt} \pa_{t} B_{(mn)}(X) = 0 \, , \label{N=2 JJL - differential constraints 1a - 4} \\
		& \pa^{m} \big\{ C_{(mn)s}(X) + C_{(sn)m}(X) - \eta_{ms} C^{a}{}_{na}(X) \big\} = 0 \, ,
	\end{align}
\end{subequations}
and the algebraic constraints
\begin{subequations} \label{N=2 JJL - algebraic constraints 1}
	\begin{align}
		& C^{m}{}_{nm}(X) = 0 \, , \\
		& \e^{rms} C_{(mn)s}(X) = 0 \, .
	\end{align}
\end{subequations}
The scaling condition \eqref{N=2 JJL - scaling constraint}, along with \eqref{N=2 JJL - algebraic constraints 1} imply that $C$ is totally symmetric, traceless and even in $X$. Following the argument presented in section \ref{subsection 4.2} we find that no such tensor exists, hence $C=0$. Furthermore, evenness in $X$ allows us to identify solutions for the remaining tensors
\begin{subequations}
	\begin{align}
		& A_{(mn)}(X) = a_{1} \frac{\eta_{mn}}{X^{3}} + a_{2} \frac{X_{m} X_{n}}{X^{5}} \, , \\[2mm]
		& B_{(mn)}(X) = b_{1} \frac{\eta_{mn}}{X^{4}} + b_{2} \frac{X_{m} X_{n}}{X^{6}} \, , \\[2mm]
		& D_{(mn)}(X) = d_{1} \frac{\eta_{mn}}{X^{5}} + d_{2} \frac{X_{m} X_{n}}{X^{7}} \, .
	\end{align}
\end{subequations}
Imposing \eqref{N=2 JJL - differential constraints 1a - 1} and \eqref{N=2 JJL - differential constraints 1a - 2} results in $a_{2} = - 3 a_{1}$, $b_{2} = -2 b_{1}$, however for this choice of coefficients \eqref{N=2 JJL - differential constraints 1a - 3} implies $B=0$, while the tensor $A$ survives. It is then easy to see that \eqref{N=2 JJL - differential constraints 1a - 4} implies $D=0$. Therefore the only solution is 
\begin{equation}
	A_{(mn)}(X) = a \, \bigg( \frac{\eta_{mn}}{X^{3}} - \frac{3 X_{m} X_{n}}{X^{5}} \bigg) \, .
\end{equation}
Now let us direct our attention to $\cH_{t}$; imposing \eqref{N=2 JJL - differential constraint 1} results in the set of equations
\begin{subequations} \label{N=2 JJL - differential constraints 1b}
	\begin{align}
		& \e^{mnt} \pa_{m} A_{t}(X) = 0 \, , \label{N=2 JJL - differential constraints 1b - 1} \\
		& \e^{mnt} \pa_{m} B_{t}(X) = 0 \, , \label{N=2 JJL - differential constraints 1b - 2} \\
		& \pa^{m} \big\{ C_{mn}(X) - \eta_{mn} C^{s}{}_{s}(X) \big\} = 0 \, , \label{N=2 JJL - differential constraints 1b - 3} \\
		& 2 \e_{qt}{}^{s} B_{s}(X) - \text{i} \pa_{t} A_{q}(X) + \text{i} \eta_{qt} \pa^{s} A_{s}(X) = 0 \, , \label{N=2 JJL - differential constraints 1b - 4} \\
		& 4 \e_{qt}{}^{s} D_{s}(X) - \text{i} \pa_{t} B_{q}(X) + \text{i} \eta_{qt} \pa^{s} B_{s}(X) = 0 \, , \label{N=2 JJL - differential constraints 1b - 5}
	\end{align}
\end{subequations}
and the algebraic constraints
\begin{subequations} \label{N=2 JJL - algebraic constraints 2}
	\begin{align}
		& \e_{n}{}^{ma} C_{ma}(X) = 0 \, , \\
		& C_{mn}(X) - \eta_{mn} C^{s}{}_{s}(X) = 0 \, .
	\end{align}
\end{subequations}
The algebraic constraints \eqref{N=2 JJL - algebraic constraints 2} imply that $C=0$. Now since $A$, $B$ and $D$ are odd in $X$ we can construct the solutions
\begin{subequations}
	\begin{align}
		& A_{t}(X) = a \frac{X_{t}}{X^{4}} \, , \\[2mm]
		& B_{t}(X) = b \frac{X_{t}}{X^{5}} \, , \\[2mm]
		& D_{t}(X) = d \frac{X_{t}}{X^{6}} \, .
	\end{align}
\end{subequations}
However it is not too difficult to show that imposing \eqref{N=2 JJL - differential constraints 1b - 4}, \eqref{N=2 JJL - differential constraints 1b - 5} requires that $A$, $B$ and $D$ must all vanish. Hence $\cH_{t}(\boldsymbol{X},\Q)=0$. 

So far we have found a single solution consistent with the supercurrent conservation equation and the point-switch identity,
\begin{align}
	\cH_{mn}(\boldsymbol{X},\Q) &= a \, \bigg( \frac{\eta_{mn}}{X^{3}} - \frac{3 X_{m} X_{n}}{X^{5}} \bigg) \, , \\[2mm]
	\cH_{\a \b \g \d}(\boldsymbol{X},\Q) &= (\g^{m})_{\a \b} (\g^{n})_{\g \d}\, \cH_{mn}(\boldsymbol{X},\Q) \nonumber \\
	&= d_{\cN=2} \, \bigg( \frac{\ve_{\a \g} \ve_{\b \d} + \ve_{\a \d} \ve_{\b \g} }{X^{3}} + \frac{3 X_{\a \b} X_{\g \d}}{X^{5}} \bigg) \, .
	\label{6.3.1}
\end{align}
Therefore the correlation function is 
\begin{equation}
	\langle J_{\a \b}(z_{1}) J_{\g \d}(z_{2}) L(z_{3}) \rangle = \frac{\boldsymbol{x}_{13 \a}{}^{\a'} \boldsymbol{x}_{13 \b}{}^{\b'} \boldsymbol{x}_{23 \g}{}^{\g'} \boldsymbol{x}_{23 \d}{}^{\d'} }{ (\boldsymbol{x}_{13}^{2})^{3} (\boldsymbol{x}_{23}^{2})^{3}} \; \cH_{\a' \b' \g' \d'}(\boldsymbol{X}_{3}, \Q_{3}) \, ,
	\label{6.3.2}
\end{equation}
where, after writing our solution in terms of the variable $\boldsymbol{X}$,
\begin{align}
	\begin{split}
		\cH_{\a \b \g \d}(\boldsymbol{X},\Q) &= d_{\cN=2} \, \bigg\{ \frac{\ve_{\a \g} \ve_{\b \d}}{\boldsymbol{X}^{3}} + \frac{\ve_{\a \d} \ve_{\b \g}}{\boldsymbol{X}^{3}} + \frac{3}{8} \frac{\ve_{\a \g} \ve_{\b \d} \Q^{4}}{\boldsymbol{X}^{5}} + \frac{3}{8} \frac{\ve_{\a \d} \ve_{\b \g} \Q^{4}}{\boldsymbol{X}^{5}} \\[2mm]
		& \hspace{30mm} + \frac{3 \boldsymbol{X}_{\a \b} \boldsymbol{X}_{\g \d}}{\boldsymbol{X}^{5}} + \frac{3 \text{i}}{2} \frac{\ve_{\a \b} \boldsymbol{X}_{\g \d} \Q^{2}}{\boldsymbol{X}^{5}} + \frac{3 \text{i}}{2} \frac{\ve_{\g \d} \boldsymbol{X}_{\a \b} \Q^{2}}{\boldsymbol{X}^{5}} \\[2mm]
		& \hspace{45mm} - \frac{3}{4} \frac{\ve_{\a \b} \ve_{\g \d} \Q^{4}}{\boldsymbol{X}^{5}} + \frac{15}{8} \frac{\boldsymbol{X}_{\a \b} \boldsymbol{X}_{\g \d} \Q^{4}}{\boldsymbol{X}^{5}} \bigg\} \, .
	\end{split}
\end{align}
However it remains to check whether this solution satisfies the flavour current conservation equation. As mentioned earlier it is difficult to check conservation laws on the third superspace point in this formalism as there are no identities that allow differential operators acting on the $z_{3}$ dependence to pass through the prefactor of \eqref{Three-point function - general ansatz}. To deal with this we will re-write our solution in terms of the three-point building block $\boldsymbol{X}_{1}$ using identities \eqref{Three-point building blocks 1a - properties 1}, \eqref{Superconformal invariants}. This ultimately has the effect
\begin{equation}
	\langle J_{\a \b}(z_{1}) J_{\g \d}(z_{2}) L(z_{3}) \rangle \hspace{3mm} \longrightarrow \hspace{3mm} \langle L(z_{3}) J_{\g \d}(z_{2}) J_{\a \b}(z_{1}) \rangle \, .
\end{equation}
Written in terms of the variable $\boldsymbol{X}_{1}$, the correlation function is found to be
\begin{equation}
	\langle L(z_{3}) J_{\g \d}(z_{2}) J_{\a \b}(z_{1}) \rangle = \frac{\boldsymbol{x}_{21 \g}{}^{\g'} \boldsymbol{x}_{21 \d}{}^{\d'} }{ \boldsymbol{x}_{31}^{2} (\boldsymbol{x}_{21}^{2})^{3}} \; \cH_{\g' \d' \a \b}(\boldsymbol{X}_{1}, \Q_{1}) \, ,
\end{equation}
where
\begin{align}
	\begin{split}
		\cH_{\g \d \a \b}(\boldsymbol{X},\Q) &= d_{\cN=2} \, \bigg\{ \frac{\boldsymbol{X}_{\g \a} \boldsymbol{X}_{\d \b}}{\boldsymbol{X}^{3}} + \frac{\boldsymbol{X}_{\g \b} \boldsymbol{X}_{\d \a}}{\boldsymbol{X}^{3}} + \frac{3}{8} \frac{\boldsymbol{X}_{\g \a} \boldsymbol{X}_{\d \b} \Q^{4}}{\boldsymbol{X}^{5}} + \frac{3}{8} \frac{\boldsymbol{X}_{\g \b} \boldsymbol{X}_{\d \a} \Q^{4}}{\boldsymbol{X}^{5}} \\[2mm]
		& \hspace{30mm} - \frac{3 \boldsymbol{X}_{\a \b} \boldsymbol{X}_{\g \d}}{\boldsymbol{X}^{3}} - \frac{3 \text{i}}{2} \frac{\ve_{\a \b} \boldsymbol{X}_{\g \d} \Q^{2}}{\boldsymbol{X}^{3}} - \frac{3 \text{i}}{2} \frac{\ve_{\g \d} \boldsymbol{X}_{\a \b} \Q^{2}}{\boldsymbol{X}^{3}} \\[2mm]
		& \hspace{45mm} + \frac{3}{4} \frac{\ve_{\a \b} \ve_{\g \d} \Q^{4}}{\boldsymbol{X}^{3}} - \frac{15}{8} \frac{\boldsymbol{X}_{\a \b} \boldsymbol{X}_{\g \d} \Q^{4}}{\boldsymbol{X}^{5}} \bigg\} \, .
	\end{split}
\end{align}
We are now able to check the conservation equation \eqref{N=2 - JJL - differential constraint 2}, which after using 
identities equivalent to \eqref{Three-point building blocks 1c - differential identities 3} becomes the constraint
\begin{equation}
	\big(\cD^{\s (I} \cD^{J)}_{\s} - \frac{1}{2} \d^{IJ} \cD^{\s K} \cD^{K}_{\s} \big) 	\cH_{\g \d \a \b}(\boldsymbol{X},\Q) = 0 \, .
\end{equation}
After a very lengthy calculation one can show that the solution above satisfies this conservation equation, 
hence this correlation function is non-trivial and is determined up to a single parameter. 

This is a peculiar result, as it was shown in section \ref{subsection 4.2} that the correlation 
function $\langle J J L \rangle$ vanishes for $\cN=1$. At first glance this appears to be a contradiction since any theory
with $\cN=2$ supersymmetry is also $\cN=1$ supersymmetric. 
However, as was discussed in the previous subsection,
the $\cN=2$ current supermultiplets $J_{\a \b}$ and $L$ contain not only the $\cN=1$ supercurrent and flavour currents,
but also the unconstrained scalar superfield $S$ and the supermultiplet of currents $S_{\a \b}$. 
Hence, non-vanishing of the $\cN=2$ three-point function~\eqref{6.3.1}, \eqref{6.3.2} implies non-vanishing of some of the three-point functions 
involving these additional $\cN=1$ currents. For example, from eqs.~\eqref{6.3.1}, \eqref{6.3.2} it follows that 
the following $\cN=1$ correlator is, in general, non-zero:
\be
\langle S_{\a_1 \a_2} (z_1) J_{\b_1\b_2 \b_3} (z_2) L_{\g} (z_3) \rangle
= - D_{(2) (\b_1}^{\mathbf{2}} D_{(3) \g}^{\mathbf{2}} 
\langle J_{\a_1 \a_2} (z_1) J_{\b_2 \b_3)} (z_2) L (z_3) \rangle |\,,
\label{6.3.3}
\ee
where the bar-projection means setting $\theta_i^{\mathbf{2}}$ to zero. In components this correlator contains (among others)
$\langle R_m (x_1) T_{pq} (x_2) V_{s} (x_3) \rangle $, where $R_{m}$ is the $U(1)$ $R$-symmetry current which exists in theories 
with $\cN=2$ supersymmetry. In theories with $\cN=1$ supersymmetry such a correlator does not exist because there is no 
$R$-symmetry current.\footnote{Note that all component three-point functions contained in~\eqref{6.3.1}, \eqref{6.3.2} 
are parity even.} On the other hand, the $\cN=2 \to \cN=1$ superspace reduction 
\be 
\langle J_{\a_1 \a_2} (z_1) J_{\b_1 \b_2} (z_2) L (z_3) \rangle \hspace{3mm} \longrightarrow \hspace{3mm}
\langle J_{\a_1 \a_2 \a_3 } (z_1) J_{\b_1 \b_2 \b_3} (z_2) L_{\g} (z_3) \rangle \, ,
\label{6.3.4}
\ee
must give zero to be consistent with the result of the previous subsection. Let us check that this is indeed the case. To perform 
the reduction we compute 
\be
- {\rm i} D_{(1) (\a_1}^{\mathbf{2}} D_{(2) (\b_1}^{\mathbf{2}} D_{(3) \g}^{\mathbf{2}} 
\langle J_{\a_2 \a_3)} (z_1) J_{\b_2 \b_3)} (z_2) L (z_3)\rangle |\,.
\label{6.3.5}
\ee
That is we must act with three covariant derivatives with respect to $\theta_i^{\mathbf{2}}$
and then set all $\theta_i^{\mathbf{2}}$ to zero. From the explicit form of the correlator 
$\langle J_{\a_1 \a_2} (z_1) J_{\b_1 \b_2} (z_2) L (z_3)\rangle$ in eqs.~\eqref{6.3.1}, \eqref{6.3.2}
it follows that it depends on $\theta_i^{\mathbf{2}} \theta_j^{\mathbf{2}} $. Since it is Grassmann even 
it contains only even powers of $\theta_i^{\mathbf{2}}$. Therefore, acting on $\langle J_{\a_1 \a_2} (z_1) J_{\b_1 \b_2} (z_2) L (z_3)\rangle$
with three derivatives as in~\eqref{6.3.5} will give a result either linear or higher order in $\theta_i^{\mathbf{2}}$, so it vanishes when we set $\theta_i^{\mathbf{2}}=0$. This shows that despite being non-zero our 
result~\eqref{6.3.1}, \eqref{6.3.2} is consistent with vanishing of the similar correlator in the $\cN=1$ case. 
 

\section{Mixed correlators in \texorpdfstring{$\cN=3,4$}{N=3,4} superconformal field theory}\label{section7}


In this section we will generalise our method for $\cN=3$ and $\cN=4$ superconformal theories. 
An essential difference with the previous cases is that the flavour current now carries $R$-symmetry indices
which must be taken into account in the irreducible decompositions.
We will start with reviewing the properties of the $\cN=3$ and $\cN=4$ supercurrent~\cite{Butter:2013goa, Kuzenko:2013vha} 
and flavour current multiplets and then apply our formalism to compute the mixed correlation functions involving these multiplets.


\subsection{Supercurrent and flavour current multiplets}


\subsubsection{\texorpdfstring{$\cN=3$}{N=3} theories}

The 3D, $\cN = 3$ supercurrent is a primary, dimension $3/2$ spinor superfield $J_{\a}$, which satisfies the conservation equation
\begin{equation}
D^{I\a} J_{\a} = 0 \, , 
\end{equation}
and has the following superconformal transformation law:
\begin{equation}
\d J_{\a} = - \x J_{\a} - \frac{3}{2} \s(z) J_{\a} + \l(z)_{\a}{}^{\b} J_{\b} \, .
\end{equation}
The two-point function is again determined up to a single real coefficient
\begin{equation}
\langle J_{\a}(z_{1}) J_{\b}(z_{2}) \rangle = b_{\cN=3} \frac{\boldsymbol{x}_{12 \a \b} }{(\boldsymbol{x}_{12}^{2})^{2}} \, .
\end{equation}
It has the right symmetry properties under permutation of superspace points
\begin{equation}
\langle J_{\a}(z_{1}) J_{\b}(z_{2}) \rangle = - \langle J_{\b}(z_{2}) J_{\a}(z_{1}) \rangle \, ,
\end{equation}
and also satisfies the conservation equation
\begin{equation}
D_{(1)}^{I \a} \langle J_{\a}(z_{1}) J_{\b}(z_{2}) \rangle = 0 \, , \hspace{10mm} z_{1} \neq z_{2} \, .
\end{equation}

The $\cN=3$ flavour current is a primary, dimension $1$ isovector $L^{I}$, which obeys the conservation equation
\begin{equation}
D^{(I}_{\a} L^{J) } - \frac{1}{3} \d^{IJ} D^{K}_{\a} L^{K} = 0 \, ,
\end{equation}
and transforms under the superconformal group as
\begin{equation}
\d L^{I} = -\x L^{I} - \s(z) L^{I} + \L^{I J}(z) L^{J} \, .
\end{equation}
The $\cN=3$ flavour current two-point function is fixed up to a single real coefficient $a_{\cN=3}$
\begin{equation}
\langle L^{I \abar}(z_{1}) L^{J \bbar}(z_{2}) \rangle = a_{\cN=3} \frac{\d^{\abar \bbar} u_{12}^{I J}}{\boldsymbol{x}_{12}^{2}} \, ,
\end{equation}
where we have introduced the flavour group index $\abar$. The two-point function obeys the correct symmetry properties under permutation of superspace points, $\langle L^{I \abar}(z_{1}) L^{J \bbar}(z_{2}) \rangle = \langle L^{J \bbar}(z_{2}) L^{I \abar}_{\a}(z_{1}) \rangle$, and also satisfies the conservation equation 
\begin{equation}
D^{(I}_{(1)\a} \langle L^{J) \abar}(z_{1}) L^{K \bbar}(z_{2}) \rangle - \frac{1}{3} \d^{IJ} D^{L}_{(1)\a} \langle L^{L \abar}(z_{1}) L^{K \bbar}(z_{2}) \rangle = 0 \, , \hspace{5mm} z_{1} \neq z_{2} \, .
\end{equation}
%


\subsubsection{\texorpdfstring{$\cN=4$}{N=4} theories}


The $\cN=4$ supercurrent is a primary, dimension $1$ scalar superfield $J$, which satisfies the conservation equation
\begin{equation}
\big( D^{I\a} D^{K}_{\a} - \frac{1}{4} \d^{I K} D^{L\a} D^{L}_{\a} \big) J = 0 \, ,  \label{N=4 - supercurrent conservation equation}
\end{equation}
and has the following superconformal transformation law:
\begin{equation}
\d J_{\a} = - \x J_{\a} - \s(z) J_{\a} \, .
\end{equation}
The dimension of the supercurrent is fixed by the conservation equation \eqref{N=4 - supercurrent conservation equation}. 
The two-point function is determined up to a single real coefficient
\begin{equation}
\langle J(z_{1}) J(z_{2}) \rangle = b_{\cN=4} \frac{1 }{\boldsymbol{x}_{12}^{2}} \, .
\end{equation}
Under permutation of superspace points, we have
\begin{equation}
\langle J(z_{1}) J(z_{2}) \rangle = \langle J(z_{2}) J(z_{1}) \rangle \, .
\end{equation}
The two-point function also satisfies the conservation equation
\begin{equation}
\big( D^{I\a}_{(1)} D^{K}_{(1)\a} - \frac{1}{4} \d^{I K} D^{L\a}_{(1)} D^{L}_{(1)\a} \big) \langle J(z_{1}) J(z_{2}) \rangle = 0 \, , \hspace{10mm} z_{1} \neq z_{2} \, .
\end{equation}

In the $\cN=4$ case there exists two inequivalent flavour current multiplets, 
described by $\sSO(4)$ bivectors $L^{IJ}_{+}$, $L^{IJ}_{-}$, which are primary with dimension 1, and satisfy
\begin{equation} \label{N=4 - flavour current symmetries}
L^{IJ}_{\pm} = - L^{JI}_{\pm} \, , \hspace{10mm} \frac{1}{2} \e^{IJKL} L^{KL}_{\pm} = \pm L^{IJ}_{\pm} \, .
\end{equation}
where $\abar$ is the index for the flavour group. The flavour current multiplets are subject to the conservation equation
\begin{equation}
D^{I}_{\a} L^{JK}_{\pm} = D^{[I}_{\a} L^{JK]}_{\pm} - \frac{2}{3} D^{L}_{\a} L^{L[J}_{\pm} \d^{K]I} \, , \label{N=4 - flavour current conservation equation}
\end{equation}
and transform under the superconformal group as
\begin{equation}
\d L^{IJ}_{\pm} = -\x L^{IJ}_{\pm} - \s(z) L^{IJ}_{\pm} + \L^{K [I}(z) L^{J]K}_{\pm} \, .
\end{equation}
Since the flavour current multiplets $L_{\pm}^{IJ}$ are inequivalent, they may be studied independently when deriving correlation functions.

\subsection{Mixed correlation functions in \texorpdfstring{$\cN=3$}{N=3} theories}

There are two mixed correlation functions in $\cN=3$ theories, they are
\begin{equation}
\langle L^{I}(z_{1}) J_{\a}(z_{2}) L^{J}(z_{3}) \rangle \, , \hspace{10mm} \langle J_{\a}(z_{1}) J_{\b}(z_{2}) L^{I}(z_{3}) \rangle \,.
\end{equation}

\subsubsection{The correlation function \texorpdfstring{$\langle L J L \rangle$}{< L J L >}}

Using the general ansatz, we have
\begin{equation}
\langle L^{I\abar}(z_{1}) J_{\a}(z_{2}) L^{J\bbar}(z_{3}) \rangle = \frac{\d^{\abar \bbar} u_{13}^{II'} \boldsymbol{x}_{23 \a}{}^{\a'} }{ \boldsymbol{x}_{13}^{2} (\boldsymbol{x}_{23}^{2})^{2}} \; \cH_{\a'}^{I'J}(\boldsymbol{X}_{3}, \Q_{3}) \, ,
\end{equation}
The correlation function is required to satisfy

\begin{enumerate}
	\item[\textbf{(i)}] \textbf{Scaling constraint:}
	
	Under scale transformations the correlation function must transform as
	\begin{equation}
	\langle L^{I\abar}(z_{1}') J_{\a}(z_{2}') L^{J\bbar}(z_{3}') \rangle = (\l^{2})^{7/2} \langle L^{I\abar}(z_{1}) J_{\a}(z_{2}) L^{J\bbar}(z_{3}) \rangle \, ,
	\end{equation}
	which gives rise to the homogeneity constraint on $\cH$:
	\begin{equation}
	\cH_{\a}^{IJ}(\l^{2} \boldsymbol{X}, \l \Q) = (\l^{2})^{-3/2} \cH_{\a}^{IJ}(\boldsymbol{X}, \Q) \, . \label{N=3 LJL - scaling constraint}
	\end{equation}

	\item[\textbf{(ii)}] \textbf{Differential constraints:}
	
	The differential constraints on the flavour current and supercurrent result in the following constraints on the correlation function:
	\begin{subequations}
		\begin{align}
		D^{(I}_{(1)\a} \langle L^{J) \abar}(z_{1}) J_{\b}(z_{2}) L^{K \bbar}(z_{3}) \rangle - \frac{1}{3} \d^{IJ} D^{L}_{(1)\a} \langle L^{L \abar}(z_{1}) J_{\b}(z_{2}) L^{K \bbar}(z_{3}) \rangle &= 0 \, ,  \\[2mm]
		D^{I\a}_{(2)} \langle L^{J\abar}(z_{1}) J_{\a}(z_{2}) L^{K\bbar}(z_{3}) \rangle = 0 \, . \hspace{20mm} &
		\end{align}
	\end{subequations}
	These equations result in the following differential constraints on $\cH$:
	\begin{subequations}
		\begin{align}
		\cD^{(I}_{\a} \cH_{\b}^{J)K}(\boldsymbol{X},\Q) - \frac{1}{3} \d^{IJ} \cD^{L}_{\a} \cH_{\b}^{LK}(\boldsymbol{X},\Q) &= 0 \, , \label{N=3 LJL - differential constraint 1} \\[2mm]
		\cQ^{I \a} \cH_{\a}^{JK}(\boldsymbol{X}, \Q) = 0 \, . \hspace{10mm} \label{N=3 LJL - differential constraint 2} &
		\end{align}
	\end{subequations}

	\item[\textbf{(iii)}] \textbf{Point permutation symmetry:}
	
	The symmetry under permutation of points ($z_{1}$ and $z_{3}$) imposes the following constraint on the correlation function:
	\begin{equation}
	\langle L^{I\abar}(z_{1}) J_{\a}(z_{2}) L^{J\bbar}(z_{3}) \rangle = \langle L^{J\bbar}(z_{3}) J_{\a}(z_{2}) L^{I\abar}(z_{1}) \rangle \, ,
	\end{equation}
	which results in the point-switch identity
	\begin{align}
	\cH_{\a}^{IJ}(\boldsymbol{X}_{3}, \Q_{3}) = -\frac{(u_{13}^{-1})^{II'} u_{13}^{JJ'} \boldsymbol{x}_{13}^{\a' \s} \boldsymbol{X}_{3 \s \s} }{\boldsymbol{X}_{3}^{4} \,  \boldsymbol{x}_{13}^{4}} \; \cH_{\a'}^{J'I'}(-\boldsymbol{X}_{1}^{\text{T}}, -\Q_{1}) \, . \label{N=3 LJL - point switch identity}
	\end{align}
\end{enumerate}

Now let's find the general solution for $\cH$ consistent with the above constraints. To do this systematically, we note that since $\cH$ is grassmann odd we must find all the linearly independent structures that are odd in $\Q$ that can be constructed out of the $\cN=3$ building blocks. A general expansion for $\cH^{IJ}_{\a}$ is
\begin{align}
\cH_{\a}^{IJ}(\boldsymbol{X}, \Q) &= \cH_{(1)\a}^{IJ}(\boldsymbol{X}, \Q) + \cH_{(3)\a}^{IJ}(\boldsymbol{X}, \Q) + \cH_{(5)\a}^{IJ}(\boldsymbol{X}, \Q) \\[2mm]
&= \cH_{(1) \a \b}(X) \, A^{IJK} \, \Q^{K \b} + \cH_{(3) \a \b \g \d}(X) \, B^{IJKLM} \, \Q^{K \b} \Q^{L \g} \Q^{M \d} \nonumber \\[2mm]
& \hspace{10mm} + \cH_{(5) \a \b \g \d \m \n}(X) \, C^{IJKLMNP}\, \Q^{K \b} \Q^{L \g} \Q^{M \d} \Q^{N \m} \Q^{P \n} \, ,
\end{align}
where $A$, $B$, $C$ are tensors formed out of the $\cN=3$ invariant tensors $\d^{IJ}$, $\e^{IJK}$. At $O(\Q^{1})$ the only choice we can make for $A$ is $A^{IJK} = \e^{IJK}$, from which we find the linearly independent structures
\begin{equation}
	\cH_{(1)\a}^{IJ}(\boldsymbol{X}, \Q) = a_{1} \, \e^{IJK} \frac{ \Q^{K}_{\a}}{X^{2}} + a_{2} \, \e^{IJK} \frac{ X_{\a \b} \Q^{K \b}}{X^{3}} \, .
\end{equation}
The conservation equation \eqref{N=3 LJL - differential constraint 2} implies that the terms $O(\Q^{1})$, are odd in $X_{\a \b}$, while the terms $O(\Q^{3})$ must be even in $X_{\a \b}$. At $O(\Q^{3})$ we have the following choices for $B$
\begin{align}
	B_{1}^{IJKLM} &= \d^{IJ} \e^{KLM} \, , \hspace{8mm} B_{2}^{IJKLM} = \e^{IJK} \d^{LM} \, ,\\[2mm]
	& B_{3}^{IJKLM} = \d^{IK} \e^{JLM} + \d^{JK} \e^{ILM} \, .
\end{align}
From which we find the linearly independent structures
\begin{align}
	\begin{split}
	\cH_{(3)\a}^{IJ}(\boldsymbol{X}, \Q) &= b_{1} \, \e^{IJK} \frac{ \Q^{K}_{\a} \Q^{2}}{X^{3}} + b_{2} \, \d^{IJ} \e^{KPQ} \Q^{K \d} \Q^{P \b} \Q^{K \g} \frac{ X_{\a (\d} X_{\b \g)}}{X^{5}} \\[2mm]
	& \hspace{25mm} + b_{3} \, ( \e^{IKP} \Q^{J \d} + \e^{JKP} \Q^{I \d} ) \, \Q^{K \b} \Q^{P \g} \frac{ X_{\a \d} X_{\b \g}}{X^{5}} \, .
	\end{split}
\end{align}
If we follow the same procedure at $O(\Q^{5})$ we find the structures
\begin{equation}
	\cH_{(5)\a}^{IJ}(\boldsymbol{X}, \Q) = c_{1} \, \e^{IJK} \frac{ \Q^{K}_{\a} \Q^{4}}{X^{5}} + c_{2} \, ( \e^{IKP} \Q^{J}_{\a} + \e^{JKP} \Q^{I}_{\a} ) \, \Q^{K \b} \Q^{P \g} \frac{ X_{\b \g} \Q^{2} }{X^{5}} \, .
\end{equation}
In determining the linearly independent terms we make use of the $\cN=3$ identity
\begin{equation}
\e^{IJK} \Q^{I \a} \Q^{J \b} \Q^{K \g} \Q^{2} = 0 \, , \label{N=3 identity 1}
\end{equation}
in addition to
\begin{subequations}
	\begin{align}
		&\Q^{I \a} \Q^{J}_{\a} \Q^{K \b} \Q^{L}_{\g} \e^{JKL}  = 2 \Q^{2} \Q^{J \b} \Q^{K}_{\g} \e^{IJK} \, , \label{N=3 identity 2} \\[2mm]
		&\Q^{I \a} \Q^{J \b} \Q^{K \g} \Q^{L \d} \e^{JKL} = - \frac{1}{2} \ve^{\a \b} \Q^{2} \Q^{J \g} \Q^{K \d} \e^{IJK} - \frac{1}{2} \ve^{\a \g} \Q^{2} \Q^{J \b} \Q^{K \d} \e^{IJK} \label{N=3 identity 3} \\[1mm]
		& \hspace{65mm} - \frac{1}{2} \ve^{\a \d} \Q^{2} \Q^{J \b} \Q^{K \g} \e^{IJK} \, , \nonumber \\[2mm]
		& \Q^{(P \a} \e^{I) M N} \Q^{M \m} \Q^{N \n} = - \frac{1}{3} \ve^{\a \m} \Q^{(P \b} \Q^{M}_{\b} \Q^{N \n} \e^{I) M N} - \frac{1}{3} \ve^{\a \n} \Q^{(P \b} \Q^{M}_{\b} \Q^{N \m} \e^{I) M N} \, , \label{N=3 identity 4}
	\end{align}
\end{subequations}
which arise as differential consequences of \eqref{N=3 identity 1}. Applying the conservation law \eqref{N=3 LJL - differential constraint 2} results in
\begin{equation}
a_{1} = b_{1} = b_{3} = c_{1} = c_{2} = 0 \, ,
\end{equation}
which leaves us with only two structures. Next we must impose the flavour current conservation equation \eqref{N=3 LJL - differential constraint 1}. 
After a lengthy calculation we find $b_{2} = \text{i} a_{2}$. Hence the solution is
\begin{align}
\cH_{\a}^{IJ}(\boldsymbol{X}, \Q) &= c_{\cN=3} \, \bigg\{ \e^{IJK} \frac{ X_{\a \b} \Q^{K \b}}{X^{3}} + \text{i} \, \d^{IJ} \e^{KPQ} \Q^{K \d} \Q^{P \s} \Q^{K \g} \frac{ X_{\a (\d} X_{\s \g)}}{X^{5}} \bigg\} \\[5mm]
&= c_{\cN=3} \, \bigg\{ \e^{IJK} \bigg( \frac{ \boldsymbol{X}_{\a \b} \Q^{K \b}}{\boldsymbol{X}^{3}} + \frac{\text{i}}{2} \frac{ \Q^{K}_{\a} \Q^{2}}{\boldsymbol{X}^{3}} + \frac{3}{8} \frac{ \boldsymbol{X}_{\a \b} \Q^{K \b} \Q^{4}}{\boldsymbol{X}^{5}} \bigg) \label{N=3 LJL - solution for H} \\[2mm]
& \hspace{45mm} + \text{i} \d^{IJ} \e^{KPQ} \Q^{K \d} \Q^{P \s} \Q^{K \g} \frac{ \boldsymbol{X}_{\a (\d} \boldsymbol{X}_{\s \g)}}{\boldsymbol{X}^{5}} \bigg\} \, , \nonumber
\end{align}
After some additional calculation it can be shown that this solution also satisfies the point-switch identity \eqref{N=3 LJL - point switch identity}, which completes our study of the $\cN=3$ correlation function.

\subsubsection{The correlation function \texorpdfstring{$\langle J J L \rangle$}{< J J L >}}

Using the general ansatz, we have
\begin{equation}
\langle J_{\a}(z_{1}) J_{\b}(z_{2}) L^{I}(z_{3}) \rangle = \frac{\boldsymbol{x}_{23 \a}{}^{\a'} \boldsymbol{x}_{23 \b}{}^{\b'} }{ (\boldsymbol{x}_{13}^{2})^{2} (\boldsymbol{x}_{23}^{2})^{2}} \; \cH_{\a' \b'}^{I}(\boldsymbol{X}_{3}, \Q_{3}) \, ,
\end{equation}
Note that in this case the flavour current is $U(1)$. The correlation function is required to satisfy

\begin{enumerate}
	\item[\textbf{(i)}] \textbf{Scaling constraint:}
	
	Under scale transformations the correlation function must transform as
	\begin{equation}
	\langle J_{\a}(z_{1}') J_{\b}(z_{2}') L^{I}(z_{3}') \rangle = (\l^{2})^{4} \langle J_{\a}(z_{1}) J_{\b}(z_{2}) L^{I}(z_{3}) \rangle \, ,
	\end{equation}
	which gives rise to the homogeneity constraint on $\cH$:
	\begin{equation}
	\cH_{\a \b}^{I}(\l^{2} \boldsymbol{X}, \l \Q) = (\l^{2})^{-2} \cH_{\a \b}^{I}(\boldsymbol{X}, \Q) \, . \label{N=3 JJL - scaling constraint}
	\end{equation}

	\item[\textbf{(ii)}] \textbf{Differential constraints:}
	
	The differential constraints on the flavour current and supercurrent result in the following constraints on the correlation function:
	\begin{subequations}
		\begin{align}
		D^{(I}_{(3)\g} \langle J_{\a}(z_{1}) J_{\b}(z_{2}) L^{J)}(z_{3}) \rangle - \frac{1}{3} \d^{IJ} D^{K}_{(3)\g} \langle J_{\a}(z_{1}) J_{\b}(z_{2}) L^{K}(z_{3}) \rangle &= 0 \, , \label{N=3 JJL - differential constraint 1} \\[2mm]
		D^{I\a}_{(1)} \langle J_{\a}(z_{1}) J_{\b}(z_{2}) L^{J}(z_{3}) \rangle = 0 \, . \hspace{20mm} \label{N=3 JJL - differential constraint 2} &
		\end{align}
	\end{subequations}
	Since \eqref{N=3 JJL - differential constraint 1} involves a covariant derivative acting on the third point, it is more difficult to impose. However it turns out that the second equation is sufficient to show that this correlation function vanishes. From \eqref{N=3 JJL - differential constraint 2} we obtain
	\begin{equation}
		\cD^{I \a} \cH_{\a \b}^{J}(\boldsymbol{X}, \Q) = 0 \, . \hspace{10mm} \label{N=3 JJL - differential constraint 2a}
	\end{equation}

	\item[\textbf{(iii)}] \textbf{Point permutation symmetry:}
	
	The symmetry under permutation of points ($z_{1}$ and $z_{2}$) imposes the following constraint on the correlation function:
	\begin{equation}
	\langle J_{\a}(z_{1}) J_{\b}(z_{2}) L^{I}(z_{3}) \rangle = - \langle J_{\b}(z_{2}) J_{\a}(z_{1}) L^{I}(z_{3}) \rangle \, ,
	\end{equation}
	which results in the point-switch identity
	\begin{align}
	\cH_{\a \b}^{I}(\boldsymbol{X}, \Q) = - \cH_{\b \a}^{I}(- \boldsymbol{X}^{\text{T}}, - \Q) \, . \label{N=3 JJL - point switch identity}
	\end{align}
\end{enumerate}

To proceed we start by decomposing $\cH$ into symmetric and anti-symmetric parts as follows
\begin{align}
	\cH^{I}_{\a \b}(\boldsymbol{X},\Q) &= \cH^{I}_{(\a \b)}(\boldsymbol{X},\Q) + \cH^{I}_{[\a \b]}(\boldsymbol{X},\Q) \nonumber \\[2mm]
	&= \cH^{I}_{(\a \b)}(\boldsymbol{X},\Q) + \ve_{\a \b} \cH^{I}(\boldsymbol{X},\Q) \, .
\end{align}
The symmetry under permutation of points \eqref{N=3 JJL - point switch identity} implies
\begin{equation}
	\cH^{I}_{(\a \b)}(\boldsymbol{X},\Q) = - \cH^{I}_{(\a \b)}(-\boldsymbol{X}^{\text{T}},-\Q) \, , \hspace{10mm} \cH^{I}(\boldsymbol{X},\Q) = \cH^{I}(-\boldsymbol{X}^{\text{T}},-\Q) \, ,
\end{equation}
therefore $\cH^{I}_{(\a \b)}$ is an odd function, while $\cH^{I}$ is an even function. They will not mix in the conservation law \eqref{N=3 JJL - differential constraint 2a}, hence we may consider each of them independently.

Starting with $\cH^{I}$, we note that since it is Grassmann even it must be an even function of $\Q$, hence it admits the expansion
\begin{align}
\begin{split}
	\cH^{I}(\boldsymbol{X},\Q) &= \cH_{(2) \a \b}(X) \, A^{IJK} \, \Q^{J \a} \Q^{K \b} + \cH_{(4) \a \b \g \d}(X) \, B^{IJKLM} \, \Q^{J \a} \Q^{K \b} \Q^{L \g} \Q^{M \d} \\[2mm]
	& \hspace{10mm} + \cH_{(6) \a \b \g \d \s \m}(X) \, C^{IJKLMNP}\, \Q^{J \a} \Q^{K \b} \Q^{L \g} \Q^{M \d} \Q^{N \s} \Q^{P \m} \, .
\end{split}
\end{align}
Here we have replaced the variable $\boldsymbol{X}$ with $X$, and introduced the arbitrary tensors $A$,$B$, and $C$, which are constructed out of the invariant tensors for the $\cN=3$ $R$-symmetry group. The $\cH_{(i)}$ are all even functions of $X$. At $O(\Q^{2})$ the only choice is $A^{IJK} = \e^{IJK}$, so we have the contribution
\begin{equation}
	\cH_{(2) (\a \b)}(X) \, \e^{IJK} \, \Q^{J \a} \Q^{K \b} \, .
\end{equation}
However, it is not too hard to see that we cannot construct an even, symmetric function of $X$ with the required index structure. Hence $\cH_{(2) (\a \b)}(X)=0$. 
Now at $O(\Q^{4})$ we have the choices
\begin{equation}
	B_{1}^{IJKLM} = \d^{IJ} \e^{KLM} \, , \hspace{8mm} B_{2}^{IJKLM} = \e^{IJK} \d^{LM} \, . \label{N=3 JJL - B tensors}
\end{equation}
The choice $B_{1}$ results in the contribution
\begin{equation}
	\cH_{(4) \a (\b \g \d)}(X) \, \Q^{I \a} \Q^{K \b} \Q^{L \g} \Q^{M \d} \e^{KLM} \, .
\end{equation}
After applying the $\cN=3$ identity \eqref{N=3 identity 2}, this is equivalent to the contribution
\begin{equation}
\cF_{(\g \d)}(X) \, \e^{IJK} \Q^{2} \Q^{J \g} \Q^{K \d} \, ,
\end{equation}
where $\cF$ is a symmetric and even function of $X$. We cannot construct such a function, hence $\cF_{(\g \d)}(X)=0$. Indeed if we follow the same procedure for $B_{2}$ we arrive at the same conclusion. Concerning contributions $O(\Q^{6})$, no terms are permitted due to the $\cN=3$ identity \eqref{N=3 identity 2}. Hence we find $\cH^{I}(\boldsymbol{X},\Q)=0$ as there are no contributions that are consistent with the algebraic symmetries.

Let us now follow the same procedure for the symmetric contribution, $\cH^{I}_{(\a \b)}$. Since it is Grassmann even it must be an even function of $\Q$, hence it must be odd in $X$. The general expansion for this contribution reads
\begin{align}
\begin{split}
\cH^{I}_{(\a \b)}(\boldsymbol{X},\Q) &= \cH_{(2) (\a \b) \m \n}(X) \, A^{IJK} \, \Q^{J \m} \Q^{K \n} \\[2mm]
& \hspace{10mm} + \cH_{(4) (\a \b) \m \n \g \d}(X) \, B^{IJKLM} \, \Q^{J \m} \Q^{K \n} \Q^{L \g} \Q^{M \d} \, ,
\end{split}
\end{align}
where the $\cH_{(i)}$ are odd functions of $X$. Here we have neglected the contribution $O(\Q^{6})$ as it will vanish due to $\cN=3$ identities. Following the same procedure outlined above we find that to $O(\Q^{2})$ we have the contribution
\begin{equation}
	\cH_{(2) (\a \b) (\m \n)}(X) \, \e^{IJK} \, \Q^{J \m} \Q^{K \n} \, .
\end{equation}
Since the tensor $\cH_{(2)}$ is symmetric in each pair of spinor indices, we may trade them for vector ones
\begin{equation}
	\cH_{(2) (\a \b) (\m \n)}(X) = (\g^{a})_{\a \b} (\g^{b})_{\m \n} \cH_{(2) a b}(X) \, .
\end{equation}
The general expansion for $\cH_{(2) a b}$ with the scaling condition \eqref{N=3 JJL - scaling constraint} is
\begin{equation}
	\cH_{(2) a b}(X) = \frac{h_{1}}{X^{3}} \eta_{a b} + \frac{h_{2}}{X^{4}} \e_{abc} X^{c} + \frac{h_{3}}{X^{5}} X_{a} X_{b} \, ,
\end{equation}
however only the second term is odd in $X$, which results in the contribution
\begin{equation}
	\cH^{I}_{(\a \b)}(\boldsymbol{X},\Q) \propto \frac{1}{X^{4}} \e^{IJK} \Q^{J \g} \Q^{K}_{(\a} X_{\b) \g} \, .
\end{equation}
Concerning the terms $O(\Q^{4})$ we follow the same procedure outlined above, for each choice of $B$ in \eqref{N=3 JJL - B tensors} we obtain the contribution
\begin{equation}
	\cH^{I}_{(\a \b)}(\boldsymbol{X},\Q) \propto \frac{1}{X^{5}} \e^{IJK} \Q^{2} \Q^{J \g} \Q^{K}_{(\a} X_{\b) \g} \, ,
\end{equation}
where we have made use of \eqref{N=3 identity 2}. Hence our solution for $\cH^{I}_{(\a \b)}$ is of the form
\begin{equation}
	\cH^{I}_{(\a \b)}(\boldsymbol{X},\Q) = \frac{a_{1}}{X^{4}} \e^{IJK} \Q^{J \g} \Q^{K}_{(\a} X_{\b) \g} + \frac{a_{2}}{X^{5}} \e^{IJK} \Q^{2} \Q^{J \g} \Q^{K}_{(\a} X_{\b) \g} \, ,
\end{equation}
It remains to impose the conservation equation \eqref{N=3 JJL - differential constraint 2a}. After a short calculation we find $a_{1} = a_{2} = 0$, hence this correlation function vanishes.

\subsection{Mixed correlation functions in \texorpdfstring{$\cN=4$}{N=4} theories}

For $\cN=4$ superconformal symmetry there are two possible mixed correlation functions (for concreteness we will 
consider only $L^{IJ}_{+}$), they are
\begin{equation}
\langle L^{IK}_{+}(z_{1}) J(z_{2} ) L^{JL}_{+}(z_{3}) \rangle \, , \hspace{10mm} \langle J(z_{1}) J(z_{2}) L^{IJ}_{+}(z_{3}) \rangle \, ,
\end{equation}
where in the second case we require a $U(1)$ flavour group. The first correlator $\langle L J L \rangle$ was previously studied in \cite{Buchbinder:2015wia}, so we will not analyse it here.

The general ansatz for the correlation function $\langle J J L \rangle$ is
\begin{equation}
\langle J(z_{1}) J(z_{2}) L^{IJ}_{+}(z_{3}) \rangle = \frac{1}{\boldsymbol{x}_{13}^{2} \boldsymbol{x}_{23}^{2}} \, \cH^{IJ}(\boldsymbol{X}_{3}, \Q_{3}) \, .
\end{equation}
As we will soon find out, the algebraic symmetries on the tensor $\cH$ are sufficient to show that this correlation function vanishes. The relevant constraints are

\begin{enumerate}
	\item[\textbf{(i)}] \textbf{Scaling constraint:}
	
	The correlation function must transform as
	\begin{equation}
	\langle J(z_{1}') J(z_{2}') L^{IJ}_{+}(z_{3}') \rangle = (\l^{2})^{3} \langle J(z_{1}) J(z_{2}) L^{IJ}_{+}(z_{3}) \rangle \, ,
	\end{equation}
	from which we find homogeneity constraint on $\cH$:
	\begin{equation}
	\cH^{IJ}(\l^{2} \boldsymbol{X} , \l \Q ) = (\l^{2})^{-1} \cH^{IJ}( \boldsymbol{X} , \Q ) . 
	\end{equation}
	
	\item[\textbf{(ii)}] \textbf{Algebraic constraints:}
	
	The symmetry under permutation of points ($z_{1}$ and $z_{2}$) constrains the correlation function as follows:
	\begin{equation}
	\langle J(z_{1}) J(z_{2}) L^{IJ}_{+}(z_{3}) \rangle = \langle J(z_{2}) J(z_{1}) L^{IJ}_{+}(z_{3}) \rangle \, ,
	\end{equation}
	which is equivalent to
	\begin{align}
	\cH^{IJ}(\boldsymbol{X} , \Q) = \cH^{IJ}( -\boldsymbol{X}^{\text{T}} , - \Q) \, . \label{N=4 JJL - Evenness}
	\end{align}
	In addition, we also have a constraints arising from anti-symmetry and self-duality of the flavour current, which give rise to
	\begin{equation}
	\cH^{IJ}(\boldsymbol{X} , \Q) = - \cH^{JI}(\boldsymbol{X} , \Q) \, , \hspace{5mm} \cH^{IJ}(\boldsymbol{X} , \Q) = \frac{1}{2} \e^{IJKL} \cH^{KL}(\boldsymbol{X} , \Q) \, . \label{N=4 JJL - antisymmetry}
	\end{equation}
	
\end{enumerate}

The constraint \eqref{N=4 JJL - Evenness} implies that $\cH^{IJ}$ is an even function, while \eqref{N=4 JJL - antisymmetry} implies that $\cH^{IJ}$ must be antisymmetric in the $R$-symmetry indices. Furthermore since $\cH$ is Grassmann even it must be an even function of $\Q$, which implies it must also be even in $X$. It is not too difficult to see that it is impossible to construct any structures consistent with these requirements out of the available building blocks, hence this correlation function must vanish.

\section*{Acknowledgements}
The authors would like to thank Jessica Hutomo, Sergei Kuzenko and Michael Ponds for valuable discussions. 
The work of E.I.B. is supported in part by the Australian 
Research Council, project No. DP200101944.
The work of B.S. is supported by the \textit{Bruce and Betty Green Postgraduate Research Scholarship} under the Australian Government Research Training Program.


\newpage

\appendix

\section{3D conventions and notation}\label{AppA}

For the Minkowski metric we use the ``mostly plus'' convention: $\eta_{mn} = \text{diag}(-1,1,1)$. Spinor indices are then raised and lowered with the $\text{SL}(2,\mathbb{R})$ invariant anti-symmetric $\varepsilon$-tensor
\begin{align}
\ve_{\a \b} = 
\begingroup
\setlength\arraycolsep{4pt}
\begin{pmatrix}
\, 0 & -1 \, \\
\, 1 & 0 \,
\end{pmatrix}
\endgroup 
\, , & \hspace{5mm}
\ve^{\a \b} =
\begingroup
\setlength\arraycolsep{4pt}
\begin{pmatrix}
\, 0 & 1 \, \\
\, -1 & 0 \,
\end{pmatrix}
\endgroup 
\, , \hspace{5mm}
\ve_{\a \g} \ve^{\g \b} = \d_{\a}{}^{\b} \, , \\[4mm]
& \hspace{-8mm} \f_{\a} = \ve_{\a \b} \, \f^{\b} \, , \hspace{10mm} \f^{\a} = \ve^{\a \b} \, \f_{\b} \, .
\end{align}
The $\g$-matrices are chosen to be real, and are expressed in terms of the Pauli matrices $\s$ as follows:
\begin{subequations}
	\begin{align}
		(\g_{0})_{\a}{}^{\b} = - \text{i} \s_{2} = 
		\begingroup
		\setlength\arraycolsep{4pt}
		\begin{pmatrix}
			\, 0 & -1 \, \\
			\, 1 & 0 \,
		\end{pmatrix}
		\endgroup 
		\, , & \hspace{8mm}
		(\g_{1})_{\a}{}^{\b} = \s_{3} = 
		\begingroup
		\setlength\arraycolsep{4pt}
		\begin{pmatrix}
			\, 1 & 0 \, \\
			\, 0 & -1 \,
		\end{pmatrix}
		\endgroup 
		\, , \\[3mm]
		(\g_{2})_{\a}{}^{\b} = - \s_{3} &= 
		\begingroup
		\setlength\arraycolsep{4pt}
		\begin{pmatrix}
			\, 0 & -1 \, \\
			\, -1 & 0 \,
		\end{pmatrix}
		\endgroup 
		\, ,
	\end{align}
\end{subequations}
\begin{equation}
	(\g_{m})_{\a \b} = \ve_{\b \d} (\g_{m})_{\a}{}^{\d} \, , \hspace{10mm} (\g_{m})^{\a \b} = \ve^{\a \d} (\g_{m})_{\d}{}^{\b} \, .
\end{equation}
The $\g$-matrices are traceless and symmetric
\begin{equation}
	(\g_{m})^{\a}{}_{\a} = 0 \, , \hspace{10mm} (\g_{m})_{\a \b} = (\g_{m})_{\b \a} \, ,
\end{equation} 
and also satisfy the Clifford algebra
\begin{equation}
	\g_{m} \g_{n} + \g_{n} \g_{m} = 2 \eta_{mn} \, .
\end{equation}
Products of $\g$-matrices are then
\begin{subequations}
	\begin{align}
		(\g_{m})_{\a}{}^{\r} (\g_{n})_{\r}{}^{\b} &= \eta_{mn} \d_{\a}{}^{\b} + \e_{mnp} (\g^{p})_{\a}{}^{\b} \, , \\[2mm]
		(\g_{m})_{\a}{}^{\r} (\g_{n})_{\r}{}^{\s} (\g_{p})_{\s}{}^{\b} &= \eta_{mn} (\g_{p})_{\a}{}^{\b} - \eta_{mp} (\g_{n})_{\a}{}^{\b} + \eta_{np} (\g_{m})_{\a}{}^{\b} + \e_{mnp} \d_{\a}{}^{\b} \, ,
	\end{align}
\end{subequations}
where we have introduced the 3D Levi-Civita tensor $\e$, with $\e^{012} = - \e_{012} = 1$. It satisfies the following identities:
\begin{subequations}
	\begin{align}
		\e_{mnp} \e_{m' n' p'} &= - \eta_{mm'} ( \eta_{nn'} \eta_{pp'} - \eta_{np'} \eta_{pn'} ) - ( n' \leftrightarrow m' ) - ( m' \leftrightarrow p' ) \, , \\
		\e_{mnp} \e^{m}{}_{ n' p'} &= - \eta_{nn'} \eta_{pp'} + \eta_{n p'} \eta_{p n'} \, , \\
		\e_{mnp} \e^{mn}{}_{ p'} &= - 2 \eta_{ p p'} \, , \\
		\e_{mnp} \e^{mnp} &= -6 \, . 
	\end{align}
\end{subequations}
We also have the orthogonality and completeness relations for the $\g$-matrices
\begin{equation}
	(\g^{m})_{\a \b} (\g_{m})^{\r \s} = - \d_{\a}{}^{\r} \d_{\b}{}^{\s}  - \d_{\a}{}^{\s}  \d_{\b}{}^{\r} \, , \hspace{5mm} (\g_{m})_{\a \b} (\g_{n})^{\a \b} = -2 \eta_{mn} \, .
\end{equation}
Finally, the $\g$-matrices are used to swap from vector to spinor indices. For example, given some three-vector $x_{m}$, it may equivalently be expressed in terms of a symmetric second-rank spinor $x_{\a \b}$ as follows:
\begin{align}
	x^{\a \b} = (\g^{m})^{\a \b} x_{m}  \, , \hspace{5mm} x_{m} = - \frac{1}{2} (\g_{m})^{\a \b} x_{\a \b} \, , \\[2mm]
	\det (x_{\a \b}) = \frac{1}{2} x^{\a \b} x_{\a \b} = - x^{m} x_{m} = -x^{2} \, .
\end{align}
The same conventions are also adopted for the spacetime partial derivatives $\partial_{m}$
\begin{align}
	\partial^{\a \b} = \partial^{m} (\g_{m})^{\a \b} \, , \hspace{5mm} \partial_{m} = - \frac{1}{2} (\g_{m})^{\a \b} \partial_{\a \b} \, , \\[2mm]
	\partial_{m} x^{n} = \d_{m}^{n} \, , \hspace{5mm} \partial_{\a \b} x^{\r \s} = - \d_{\a}{}^{\r} \d_{\b}{}^{\s}  - \d_{\a}{}^{\s}  \d_{\b}{}^{\r} \, ,
\end{align}
\begin{equation}
	\x^{m} \partial_{m} = - \frac{1}{2} \x^{\a \b} \partial_{\a \b} \, .
\end{equation}
We also define the supersymmetry generators $Q^{I}_{\a}$
\begin{equation}
	Q^{I}_{\a} = \text{i} \frac{\partial}{\partial \q^{\a}_{I}} + (\g^{m})_{\a \b} \q^{I \b} \frac{\partial}{\partial x^{m}} \, , \label{Supercharges}
\end{equation}
and the covariant spinor derivatives
\begin{equation}
	D^{I}_{\a} = \frac{\partial}{\partial \q^{\a}_{I}} + \text{i} (\g^{m})_{\a \b} \q^{I \b} \frac{\partial}{\partial x^{m}} \, , \label{Covariant spinor derivatives}
\end{equation}
which anti-commute with the supersymmetry generators, $\{ Q^{I}_{\a} , D^{J}_{\b}\} = 0$, and obey the standard anti-commutation relations
\begin{equation}
	\big\{ D^{I}_{\a} , D^{J}_{\b} \, \big\} = 2 \text{i} \, \d^{IJ} (\g^{m})_{\a \b} \partial_{m} \, .
\end{equation}
	

\printbibliography[heading=bibintoc,title={References}]


\end{document}